\documentclass[10pt]{article}

\usepackage{amsmath}
\usepackage{amssymb}

\usepackage{graphicx}

\usepackage{cite}

\usepackage{color} 

\usepackage[labelfont=bf,labelsep=period,justification=raggedright]{caption}

\makeatletter
\renewcommand{\@biblabel}[1]{\quad#1.}
\makeatother

\date{}

\begin{document}

\begin{flushleft}
{\Large
\textbf{Bio-inspired Mechanism and Model Exploration of Software Aging}
}
\\
Pengfei Chen$^{1,\ast}$, 
Yong Qi$^{1}$, 
Di Hou$^{1}$, 
Jiankang Liu$^{2}$
\\
\bf{1} Computer Science and Egineer, Xi'an jiaotong university, Xi'an, Sha'an Xi,China
\\
\bf{2} School of Life Science and Technology, Xi'an jiaotong university, Xi'an, Sha'an Xi,China
$\ast$ E-mail: chenpengfei@outlook.com
\end{flushleft}

\section*{Abstract}
Software systems situated in network environment may experience performance degradation, availability decrease and even crash during long time running, which is called software aging. This phenomenon has been studied for more than 15 years, but most of the literatures studied software as a black box, none of them uncovered the fundamental and widely accepted mechanism of software aging as far as we know. Through analyzing the characteristics between biological aging and software aging, we find some interesting common points and bridge the gap between these two seemingly unrelated phenomena. The free radical aging theory in biological studies is also applicative to explore the mechanism and model of software aging. This paper finds an equivalent concept named `software free radical' in software aging to free radical in biological aging. In our study, the accumulation of `software free radical' is a root cause of software aging. Using the free radical modeling methodology in biological aging, we give a model for describing the kinetic of software aging based on feedback loops. Although this paper doesn't give enough theoretical proof of the modeling method, the practical results show that the feedback loop model can
describe the kinetic of software aging precisely. To further validate the aging mechanism, we propose several software rejuvenation strategies focusing on cleaning the `software free radical'. The results show that software aging can be mitigated effectively by strengthening negative feedback loop or weakening positive feedback loop. This paper is the first try to answer the question `How software ages' through interdisciplinary studies. Leveraging the conclusions in this paper, people can design better software systems or keep their systems at a high performance level during long time running.
\section*{Author Summary}
Although a large body of literatures have studied software aging, the mechanism of software aging still
remains uncovered. Due to the multiple endogenous and exogenous impact factors, software aging is
a complex stochastic process. Considering the similarities between software and organism, we propose
a novel bio-inspired approach to explore the mechanism and model of software aging. We bridge the
gap between software aging and biological aging using free radical aging theory and adopt the positive-
negative feedback loop model to describe the kinetic of software aging. Through experimental evaluation
in several open source software systems, we conclude the `software free radical' aging viewpoint can
explain the mechanism of software aging well and the feedback loop model can describe the kinetic of
software aging precisely. As far as we know, this is the first work to study software aging using bio-inspired
approaches.
\section*{Introduction}
As we all know, human beings or other organisms age with the passage of time. However very few people
know aging phenomenon also exits in software systems. In the mid-90s of last century, the concept of
`software aging' is proposed and depicted as the phenomenon of system degradation including the increase
of system failure rate and the decrease of performance \cite{1} . In some mission-critical system, software aging
not only causes revenue loss \cite{3,7,10} but also human lives \cite{2}. Through more than a decade's research, a mass
of models and methods are proposed to study software aging from several aspects including the aging
reasons \cite{16,17,18,19}, aging models \cite{5,7,11,14,20,23}, anti-aging strategies (i.e. software rejuvenation) \cite{1,3,4,5,6,9,12,13,15}  and so on. But the fundamental mechanism of software
aging is still a mystery. This dilemma hinders a further understanding of software systems and derives
some inefficient anti-aging strategies. Focusing on this problem, we make a first step to explore the
underlying mechanism of software aging through interdisciplinary studies between biology and software. \\
\textbf{Software aging and rejuvenation}

The software aging phenomenon was first proposed by Huang in 1995 \cite{1}. A widely accepted definition
of software aging is: `As the running time increasing, the software system manifests performance decrease, service quality degradation and even system
crash' \cite{1}. Although this concept is still controversial, it has been observed in many real systems such
as Linux operating system \cite{20,13} , Java virtual machine \cite{26}, Web server \cite{7,14}, etc. Memory leak, unreleased file
lock, round-off error and data corruption are known partial causes of software aging. But a unified and
fundamental mechanism of software aging to answer the question `why and how software aging happens'
is still debated. Grottke \cite{16,17,18}, proposed aging-related bug was an important reason to software aging. This
kind of bug accompanies the system for a long period and exhibits its harm occasionally at runtime.
Although it is a reasonable explanation of software aging, the detection and influence measurement of
these bugs are extraordinarily hard. Therefore it's difficult to use this concept to help us to counteract software
aging at runtime. Moreover, the increasing complexity in vertical level (i.e. code base increases) and
horizontal level (i.e. more distributed) makes the exploration of hidden aging-related bug more difficult.

To counteract software aging, the concept of software rejuvenation was also proposed in \cite{1}. It is
informally defined as a process of recovering software systems through proactively cleaning the internal
state of the system at runtime. However lack of the support of mechanism of software aging, the current
rejuvenation strategies \cite{4,5}(e.g. hard reboot) achieve the recovery at the expense of high cost (e.g. long
downtime). Considering the dilemma of currant research, it's necessary to seek more practical mechanism
of software aging.

In order to understand software aging, lots of researchers devoted themselves to this work. They built
many different mathematical models to describe aging process and carry out software rejuvenation in
different software systems. The main research methods can be divided into model-based methods and
measurement-based methods. And the primary contents include software aging modeling, software aging
predication, software aging pattern recognition, software aging root cause analysis and optimal software
rejuvenation scheduling. The goal of model-based methods is either to analyze system availability at
steady state or transient state or to obtain optimal rejuvenation schedule strategies taking maximum
availability or minimum rejuvenation cost as targets with Markov model, queuing model or other math-
ematical model. Huang .el \cite{1} first analyzed the aging of telecom system using three-state Markov model
and presented a simple rejuvenation scheduling strategy . Dohi improved Huang's Makov model in his
literatures \cite{10,13}. He proposed a new analytical model called Semi-Markov model. Trivedi did many semi-
nal work on software aging model analysis and plenty of research emerged following his work. He first
introduced stochastic reward network to software aging study \cite{4}. And he advanced multi-state analysis
in transaction based applications \cite{5}. Instead of stochastic model, Andrzejak \cite{21} proposed a deterministic
model to software aging and software rejuvenation assuming system degrades depending on the work
done by last rejuvenation. Recently a new rejuvenation method based on dynamic pro-
gramming such as MDP (Markov Decision Process) \cite{27} and Q-learning \cite{28}  were adopted. These methods
could handle more complex rejuvenation model with a great many states compared to previous analytical
method. The measurement-based methods need to monitor all kinds of system performance counters
such as CPU utilization, surplus of physical memory and so on and estimate aging speed, aging time
point, aging probability or software rejuvenation time. Garg \cite{23} put forward a method to evaluate the
state of software aging based on measurement for the first time. In his literature he first collected some
and idle communication channel and then evaluated the state of software aging using these data. L.Li
.etl \cite{14} proposed a method to predicate resource consumption based on time-series analysis. He correlated
software aging and resource consumption with time-series analysis method. Then he analyzed the data
collected from apache web server with linear-regression and concluded software aging existed in apache
web server. Combining the model-based methods and measurement-based methods, Vaidyanathan, etl \cite{24}
classified the system workload data into eight state with cluster method and modeled the workload states
using a Semi-Markov process. If a workload state was given, his model could calculate the swap space
and memory consumption. And he proved workload was an important factor of software aging. Different
from previous analytical methods, Andrzejak \cite{25} used machine learning to predicate software aging in
SOA (Service-Oriented Architecture) application. Considering a seasonal phenomenon in web system
aging, Grottke modified traditional time series method to fit seasonal pattern in his paper \cite{7}. In order
to conduct software aging experiments, Matias \cite{22} presented an experiment methodology to accelerate
software aging and used a parametric model to estimate the lifetime of the investigated system. \\
\textbf{Biological aging}

Biological aging has been studied for hundreds of years. More than 300 kinds of aging theories
are proposed \cite{29}. All of these theories can be divided into two kinds: biochemistry theory and evolution theory.
The former theory mainly cares about the progressive accumulation of materials in cells, tissues and organism. Free radical methylation and telomere reduction are also included in this theory. The molecule biological experiments are the foundation of this theory. The evolution theory  uses
reproduction rate, mutation, inheritance and natural selection to explain aging. Here we don't want to
know why aging, we just take into account how aging happens. Therefore, our focus is on biochemistry
theory. Actually, this theory can be divided into three classes according to whether randomness is considered \cite{30}. The first class is simple deterioration. It describes aging is simply the accumulative result of universal deteriorative process such as oxidation, molecular damage, wear and tear, or accumulation of
adverse byproducts. The second class is no-programmed aging also named passive aging. It describes aging as a passive result of an organism's inability to resist fundamental deteriorative process. That means the biology cannot adapt to the damage caused by the outer world. The last class is programmed
aging. It is an active aging process describing aging is an adaptation and purposeful design feature. According to this theory, the life span of biology is designed by genes. All of
these theories have strength and weakness. They can explain some aging phenomenon but can't explain
others. Strictly speaking, they are not theories, but hypotheses. Aging is a very complex process and a
single theory can hardly describe aging completely. So in this paper a model based on free radial aging
is built combing these three theories.

Free radicals are normally produced in endogenous metabolic reactions. They are atoms, molecules,
or ions with unpaired electrons and highly chemically reactive. A small amount of free radicals are
beneficial in biology. But over dose can cause damages to biology such as gene mutations \cite{29,31,32}, abnormal
protein metabolism \cite{34} and so on. A regulation mechanism of free radical in biology keeps free radicals
at a normal level. But as the age of biology increasing, free radicals progressively accumulate in biology
and finally the biology goes to die. In this theory, the free radical production process and regulation
process in organism are dominated by genes. So this theory includes programming aging factor. And the
adaptation to damages caused by over dose free radicals can be considered as a non-programmed aging
factor. The aging process is also a free radical accumulation process, so a simple deterioration is also
included in this theory. Hence free radical aging process has overlap with those three aging theory mentioned
above. And this theory is widely discussed. If something like free radical could be found in software systems,
this theory could be used to explain software aging mechanism and some models in this field could be
used to model software aging.

Comparing to a short history of software aging, biological aging has been studied for hundreds of
years. Thousands of papers were published to discuss biological aging. Some papers put their emphasis
on phenotypes of aging and others were more concerned with mechanism of aging. Stadlbauer \cite{35} studied
the aging-related changes in central nervous system. He concluded quantitative fiber tracking enables
identification of differences in diffusivity and fiber characteristics due to normal aging. David \cite{36} used
structural and functional neuroimaging techniques to investigate the effect of aging to human cognitive function.
Marineo \cite{40},Silva \cite{38}, Hershey \cite{39} and Luo \cite{37} proposed entropy increasing was a important
phenotype of biological aging from the point of thermodynamics. Researchers have conducted many
mechanism studies at different granularity such as molecule level, cell level, tissue level, organ systems and so on. Here we will put
our emphasis on the studies at molecule level. \\
\textbf{Analogies between software and organism}

According to our investigation, Sergei et al \cite{41} have proposed that surprising similarities exist between
biological and technological systems. They proved their perspective through comparing the distribution of component frequency of utilization between these two systems. Organism is coded by the underlying genes.
Genes hold the information to build and maintain the organization of organism and control the whole
life process in the body of organism. A most important work of genes is to control the synthetization
of proteins through translation and transcription. Proteins are the material foundation of life. They
are the basic functional units in organism. Organism needs materials and energy to keep alive. These
materials and energy are generated or transformed in the complex metabolic processes. Based on the
metabolic processes, organism manifests all kinds of life phenomenon such as growth, aging and death.
Actually software systems have the similar structures and processes as organism. Here we bridge some
concepts between software and organism. `Software genes' are some segments of source code. AS the
same function of biological genes, software genes hold all the information of software systems. They take
the responsibility of regulating the execution paths and determining when to start or stop software. In
real software systems, these software genes form lots of components undertaking different functions such
as transmitting data from disk to memory, scheduling processes or threads and else. These functional
components are most like proteins in organism. The main task of software systems is to process information. The processes that software systems utilize system resources including CPU, disk, memory and
network to process information are recognized as  the 'metabolic processes' of software systems. Differently, in software
metabolic processes, the reactants and productions are resources and data rather than  materials and energy .
Built on the `metabolic process', software system may also have some similar life phenomena to organism. Therefore we have reasons to believe there may be some common points between software aging and
biological aging. And these common points may be the potential keys to unlock the Pandora's Box of
software aging. Figure 1 demonstrates the similarities between software and organism and the conjecture
of some common theories and methods existing between software aging and biological aging. \\
\textbf{Research contributions}

In this paper, we explore the similarities between organism and software qualitatively and build a
similar concept stack between these two entities. Driven by these similarities, we find that the free
radical aging theory in biological area can also be applied in software aging. An equivalent concept
named `software free radical' (abbreviated as SFR) to `biological free radical' (abbreviated as BFR) is
established. Using the free radical aging theory, we can explain the fundamental mechanism of software
aging. According to the SFR generation and subduction process during software running period, we
give an aging model based on feedback loops. To validate the proposed aging mechanism and model, we
conduct several full-fledged experiments in three well-known open source software systems. The results
show that feedback loop model can precisely describe the kinetic of software aging. In a streaming media
software system named Helix Server, we conduct a complete analysis from source code and find the
SFR generation and subduction processes. Finally we propose several software rejuvenation strategies
through strengthening the SFR subduction process (i.e. negative feedback loop) and weakening the SFR
generation process (i.e. positive feedback loop). The experiment results show the good effectiveness of
these rejuvenation strategies. Please note that, we just provide a complementary view from an interdisciplinary study rather than  disprove the existing conclusions.
\section*{Results}
\subsection*{Mechanism of software aging}
In the introduction section, we have introduced what's the software aging, what's the biological aging, what's
the free radical aging theory and what's the relationship between software and organism. In this section, we
will introduce the mechanism of software aging according to free radical aging theory, the results of the
aging model based on feedback loops and the effectiveness of several rejuvenation strategies. \\
\textbf{Analogy between software aging and biological aging} 

The relationship between software and organism has been discussed in the introduction section. Motivated by the similar structures and processes between these two entities, we further discover the following
similarities between software aging and biological aging. In macro level, four common points are listed
here:
\begin{itemize}
\item Universality. Software aging is a universal phenomenon in software systems just as mentioned in
the introduction section while biological aging is also universal in organisms.
\item Accumulation. Software aging and biological aging are both progressive accumulative process as
time goes by.
\item Complexity. Software systems are consisted of all kinds of components interacting with the complex
network environment. Biology is made up of all kinds of cells and lives in more complex natural
environment. From the point of system theory, both of them belong to complex systems. These two
aging phenomena are the holistic expression of complex interactions among different components.
\item Uncertainty. Software aging and biological aging are both uncertain with many undetermined
factors involved. Hence two processes are more often than not modeled by stochastic models.
\end{itemize}
In micro level, there are three common points between software aging and biological aging:
\begin{itemize}
\item Genes defects. Some literatures \cite{29,31,32} point out gene mutations may occur during organism aging.
Similarly source code of some software systems may be damaged because of stack/data corruption
in the aging process causing system crash.
\item Protein changes. In biological aging process the proteins may deceases or lose functionalities \cite{34}
while in software aging process the software components may not work or work with errors, so some
exceptions or error messages are thrown.
\item Metabolic chaos. Normal materials and energy metabolism are critical in organism. But during
organism aging, the metabolism may become chaotic which means the normal processes are violated,
several studies \cite{37,38,39,40} support this perspective. There is the same phenomenon in software aging process, \cite{42,43}
showed that the memory resource consumption exhibited a complexity increasing characteristic.
\end{itemize}
These common points between software aging and biological aging in macro and micro level further
strengthen our belief in taking advantage of biological aging theories to study software aging. We
find free radical aging can be used to describe and model software aging mechanism. In the following we will explain
why and how to choose free radical aging to analyze software aging mechanism. \\
\textbf{Metabolism of SFR in software systems}

Before demonstrating the mechanism of software aging, we first illustrate the basic mechanism of free
radical aging theory in biological studies. Free radicals are normally produced in endogenous metabolic
reactions. The procedures of free radical generation and damage are demonstrated in Figure 2. We strongly recommend the reader to refer to this paper \cite{44} for details. Generally speaking, free radicals are generated in the materials and energy metabolic reaction procedures proceeding 
in mitochondria. Free radicals result in organism aging in three ways. First they can cause mtDNA
mutation leading to metabolism abnormal behavior. Second they affect the biosynthesis process through decreasing
the amount of proteins. And last free radicals oxidize proteins causing cell apoptosis. The procedures
mentioned above are called free radical `vicious cycle' \cite{44} forming a positive feedback loop. But in the body
of organism, a mechanism which could sense and adjust the level of free radical exists. This mechanism
could be called `virtuous cycle' forming a negative feedback loop. Figure 3 integrates the positive feedback
loop and negative feedback loop. The detailed description of these two feedbacks is presented in literature \cite{44}.

Then how should we define `software free radical'? During the running processes of software systems,
something like free radical is generated indeed. Briefly speaking, software free radical is defined as \textbf{data
objects which are useless in the current given computation context}. Alike the classification of BFR, we divide
SFR into four classes. Data objects belonging to one or more of the following categories are defined as
software free radical.
\begin{itemize}
\item Used data. Some data objects that have been used to finish their services and then become useless
to the users, such as cache data or leaked data.
\item Unwanted data. The intermediate data objects such as temp data generated during software running.
\item Malicious data. Data objects poured into software systems by malicious programs.
\item Corrupted data. Some incorrect data or damaged data during software running.
\end{itemize}
Here we don't consider SFR due to malicious attacks, but the one generated in the normal software
metabolic processes. A simple reason is that the malicious data could be avoided if detected and managed
appropriately. However, other SFR are hard to be avoided. The information processing of software
systems begins with input data such as transaction request. Then they schedule resources such as CPU,
memory, network or disk to process the input data and produces output data. Simultaneously, some
resources are released. These released resources are used again in the next computation cycle. Software
free radicals are generated in the process of information processing. They can decrease the efficiency of
some parts of software system, corrupt program, or exhaust resources which speed up the generation of
software free radicals. So finally the software system gets slower and slower and finally crashes down. The
details of these processes are demonstrated in Figure 4. Similar to free radical `vicious cycle' in organism,
we also define those processes mentioned above as software free radical `vicious cycle' forming positive
feedback loop in software system. However most of software systems have the ability to clear software
free radicals. Garbage collection in JVM (java virtual machine), memory releasing in applications or
some other adaptive adjusting mechanisms are called `virtuous cycle' forming the negative feedback loop.
It mitigates the software aging process. The upper part of Figure 5 presents software free radical positive
feedback loop and the lower part presents the negative feedback loop.

From analysis above, we can see the accumulating process of software free radical integrates a negative
feedback loop and a positive feedback loop. And this is also the software aging process. In the rest of
this paper, we don't distinguish software aging process and software free radical accumulation process
any more.
\subsection*{Aging trend}
In this paper, we conducted a large quantity of experiments on three well-known open source software
platforms including web server (Apache Httpd), a stream media server (Helix server) and a transaction
processing system (TPC-W). The aging phenomenon of web server has been observed and studied in
previous work \cite{7,14}. In the following part we first use the aging model based on feedback loops to describe
the kinetic of the three open source software aging procedure and then put the emphasis on the Helix server
system to analyze the mechanism of software aging and the effectiveness of several software rejuvenation
strategies.

For Helix server, we only select three representative results under different workload types due to the limited space, denoted as
$w_{1}$,$w_{2}$,$w_{3}$.The bandwidth per player generated by workload $w_{1}$,$w_{2}$,$w_{3}$ are demonstrated from Figure 6 to
Figure 8 and the working set are demonstrated from Figure 9 to Figure 11.

Even a cursory glance at these figures, a decreasing trend in \textit{bandwidth} and an increasing trend in
\textit{working set} are observed during software running. Due to the noise involved in the raw data, we use
\textbf{Lowess} \cite{45} method to smooth these time series. The parameter $f$ (i.e. the proportion taken by the data
points joining in the local regression in the total data points when using \textbf{Lowess}) should be determined first.
Four trials are conducted in \textit{bandwidth} smoothing and \textit{working set} smoothing shown in Figure 12 and
Figure 13. $f = 0.3$ is the tradeoff between computation efficiency and precision. Taking $f = 0.3$, other
\textit{bandwidth} and \textit{working set} generated by $w_{1}$ and $w_{3}$, are exhibited with \textbf{Lowess} smoothing in Figure
14 and Figure 15. From Figure 12 to Figure 15, we can observe that: at the initial phase of the
software system, the aging speed is slow; after a sharp increasing phase, the aging speed tends to be
zero, in this situation the system can't provide normal service, because if the bandwidth is less than 30 /Kbyte, serious video and audio frame loss are generated . The mechanism of streaming system aging and quantitative model will be given in next part.
For web server, the response time of web pages is demonstrated in Figure 16. After using \textbf{Lowess}, $f =
0.3$ to smooth the original data and achieve the results shown in Figure 17. From these two figures, an
increasing trend can be observed clearly.
For TPC-W, we conducted several experiments repeatedly. After a few hours' running, MySQL service
stops responding and finally crashes. We recorded the memory consumption of MySQL during the whole
procedure which is shown in Figure 18 and the smoothed version is shown in Figure 19.
\subsection*{Model verification}
In our experiments, memory consumption, bandwidth per player and web page response time all can be
used as indicators of software aging except a little difference that is a lower bandwidth per player indicates a 
higher aging degree while a higher memory consumption or web page response time indicates a higher aging
degree. Before model verification, the raw data should be preprocessed appropriately. Moreover due to the
different scales of these indicators, we normalize the raw data to the range (0,1). As our aging model just
describes the kinetic or the trend of software aging rather than the detailed changes, we use \textbf{Lowess}
to smooth the little changes embedded in the raw data. The transformation procedure is:
For 'bandwidth per player' indicator:
\begin{equation}
Y(t)=\frac{max(Lowess(I(t)))-Lowess(I(t))}{max(Lowess(I(t)))-min(Lowess(I(t)))}
\end{equation}
For memory consumption or web page response time:
\begin{equation}
Y(t)=\frac{Lowess(I(t))-min(Lowess(I(t)))}{max(Lowess(I(t)))-min(Lowess(I(t)))}
\end{equation}
Here $Y (t)$ denotes the degree of software aging and $I(t)$ denotes the indicator data series. In addition,
we changed the temporal scale from second to hour as the model is an exponential function of time. If
not, the coefficients of our model are too small to be estimated. Using \textit{maximum likelihood} estimation
method, we find out the coefficients of the aging model in different data series. And two standard metrics
named RMSE (i.e. Root of Mean Square Error) and R-square are leveraged to value the precision of our
model. RMSE is a measure of the differences between values predicted by a model or an estimator and the values actually observed defined as follows:
\begin{equation}
RMSE=\sqrt{\frac{\sum_{t=1}^{n}({\hat{I}}_{t}-{I}_{t})}{n}}
\end{equation}
Where ${\hat{I}}_{t}$ is the estimated or predicated value at time $t$ and ${I}_{t}$ is the observed value at time $t$. R-square
provides a measure of how well observed outcomes are replicated by the model, as the proportion of total 
variation of outcomes explained by the model []. It is defined as: 
\begin{equation}
R^{2}=1-\frac{SS_{res}}{SS_{tot}}
\end{equation}
where  $$SS_{res}=\sum_{t=1}^{n}(I_{t}-\bar{I})^{2},SS_{tot}=\sum_{t=1}^{n}(I_{t}-\hat{I}_{t})^{2}$$
Here $SS_{res}$ means sum of squares of residuals, $SS_{tot}$ means sum of squares of total, $I_{t}$ means the observed
value at time $t$, $\bar{I}$ means the average of the observed value and  $\hat{I}_{t}$ means the predicted or estimated value
at time $t$. The lower of RMSE or the higher $R^{2}$ implies a better model.

Table 1 demonstrates all the estimated model parameters $K$, $\alpha$, $\beta$ and RMSE, $R^{2}$. And the fitting
results using our model are listed from Figure 20 to Figure 24. From these results we can easily see all the
RMSE are lower than 0.08 and all the $R^{2}$ higher than 0.93 tending to 1 which means the aging model
based on feedback loops can describe the kinetic of software aging appropriately. According to these
results, it's easy to understand or predict the dynamic variation of software aging of different software
systems. What a prerequisite we need is to estimate the three parameters of this model.
\subsection*{Software aging mechanism of Helix server}
Similar to the complex metabolic network in organism body, complex metabolic processes are proceeding
in software systems. In these processes, SFR is generated. In Helix server, we find this free radical, namely
the media content cached by the system and other relative data piled in memory. Most of streaming
system implements cache to buffer part of media contents in order to accelerate data delivering. When
users request media files, the streaming system first fetches the content from memory cache and if not
got, fetches from the disk.

When the request number is low, the request arriving interval is long or the request files are always
the hot objects (i.e. popular media files which are accessed most), the cache mechanism is extraordinarily
effective. The memory resource will not be over exhausted as the cached content is released in time.
But when the adverse situation happens, some cache content couldn't be released and the memory
resource decreases gradually. In order to describe this cumulative process, two experiments with two
different workloads are conducted. One workload is (600, 0, 20, 20, 1000, 0) denoted by $L_{1}$ and the other
is (600, 0, 100, 20, 1000, 0), denoted by $L_{2}$ . There is only one difference between $L_{1}$  and $L_{2}$  that $L_{1}$  can
only access the fixed 20 files within 100 files but $L_{2}$  has authorization to access all the 100 files. In
other words these two workloads have different request file distribution. When Helix server runs under
L1, there is no significant increase in cache content (See the black line in Figure 25), shaking around
100MB. And the memory is allocated and freed periodically with very tiny increase (See the black line in Figure 26, the
memory consumption is periodic as our workload is periodic.). The bandwidth per player keeps around
120kbyte (See the black line in Figure 27). But when Helix server runs under $L_{2}$ , the cache content increases sharply from
80MB to around 200MB (See the blue line in Figure 25). The utilized memory from 600MB to 1100MB
(See the blue line in Figure 26) and the bandwidth decreases from 120kbyte to around 30kbyte which is lower than the
threshold supporting normal streaming service (See the blue line in Figure 27). These accumulated data
objects in memory causing performance degradation are SFR.

In a real streaming service, workload from outside is very complex and variable in the manner of
request frequency and the size of request files. The software system can hardly clean up the cached
content in time resulting in SFR accumulation inevitably. Similar to the feedback loops in organism,
there are negative and positive feedbacks in Helix server. Through analyzing the source code of Helix
server, we find these two feedbacks indeed. As cached content increasing, the available memory resource
deceases. This leads to transportation time per read from disk increasing. If this time exceeds data
package processing time, Helix server increases the size of data block per read from disk until to the
maximum 16Kb. This aggravates disk burden causing disk queue increasing sharply (See Figure 28).
Lots of requests waiting in the queue further increase the cached content. And consequently time per
read and the size of data block per read increases again. In this situation, the subsequent requests can't
be served and the transportation bandwidth becomes low. This is the positive feedback loop of SFR in
Helix server. However some memory cleaning mechanisms have been built in Helix server. One thread
of Helix server names `memreaper' would clean the memory blocks which have not been accessed for a
long time. And the operating system is able to reclaim some of data objects freed by Helix server. These
mechanisms make up the negative feedback loop in Helix server. Figure 29 presents the positive and
negative feedback loops in Helix server, here the character ``+" denotes the positive impact between two
successive components and ``-" denotes the negative impact. The throughput of Helix server changes
with the accumulation of SFR. These two feedbacks are cornerstone of our idea. To further validate
the mechanism of software aging proposed in this paper, we conducted several software rejuvenation
strategies focusing on cleaning up or mitigating the SFR.
\subsection*{Software rejuvenation}
From analysis above, we know software free radical is a very important factor causing software aging and
the aging speed is dominated by the negative and positive feedback loops. So if we reduce the amount
of software free radical, the software system may recovery from software aging partially or thoroughly.
Hard reboot is a most effective rejuvenation method. It can sweep out software free radicals in the
memory thoroughly, but it always causes service interruption which is not allowed in some safety-critical or
mission-critical systems. Hence reboot is not considered in this paper. Considering the generation process
of software free radical, this paper introduces several new rejuvenation strategies including strengthening
the negative feedback loop and weakening the positive feedback loop. Two methods are used to weaken
the positive feedback loop. They are load regulation and disk queue regulation. The former method
doesn't need to modify source code of the software system. But the latter method needs to modify source
code of the software system through adding automatic disk queue adjusting algorithm. One method is
implemented to strengthen the negative feedback loop. It is to modify the memory cleaning mechanism
in the source code of Helix server. Our rejuvenation methods are triggered when the aging degree has
reached the threshold preset by users. \\
\textbf{Rejuvenation based on workload regulation}

In Helix server, workload is a factor influencing the generation of software free radical. So if the
workload is regulated appropriately, the positive feedback loop will be weakened and software aging is mitigated. We
have designed two workload regulation methods:
\begin{itemize}
\item If the file accessed by a new request has been cached in the memory, the request will be accepted
otherwise it will be migrated to another server.
\item New requests are accepted in a specific probability, the rejected requests are migrated to other
servers.
\end{itemize}
New requests are accepted in a specific probability, the rejected requests are migrated to other
servers.
One experiment implements the former rejuvenation method
and the other implements the latter rejuvenation method. The bandwidth per player, memory consumption and disk queue length are observed. Taking method 1, the results are presented from Figure 30 to 32 (\textbf{Remark}: the red line denotes data series before rejuvenation and the blue line denotes data series
after rejuvenation, it's the same in other rejuvenation experiments).
From the three Figures we can see at the initial phase of Helix server the bandwidth stays at a high
level, but after a period of time it decreases below 40kbytes, at this time the system is out of service.
At 2000x15 seconds point, rejuvenation method 1 is conducted. The bandwidth increases significantly,
at the same time the disk queue decreases significantly and the memory exhaustion decreases slightly.
Helix server recoveries from aging state partially, so this method is effective. However it doesn't clean
the software free radical and just reduces the generation speed of software free radical. Therefore, the
memory consumption changes slightly. Taking method 2, the results are presented in Figure 33-35.In
this experiment, e similar aging process of Helix server is observed. At 2500x15 seconds time point,
rejuvenation method 2 is conducted. After the rejuvenation, the bandwidth increases sharply, the disk
queue and memory exhaustion decreases significantly too. Intuitively observed, rejuvenation method
2 is better than method 1. But method 2 confronts a hard problem: how to set the probability. If
the probability is set too high, the rejuvenation effect is not significant. But if it is set too low, the
computation resource is wasted. So the probability should be set appropriately. A deep discussion about
this problem will be discussed in our future work. \\
\textbf{Rejuvenation based on disk queue regulation}

In Helix server, the factor of disk queue is in the positive feedback loop. Disk queue length increasing
will cause software free radical accumulation. We have added a thread in the original Helix server source
code. When the service quality (i.e. bandwidth per player) decreases and the size of data block per read
from disk is more than 4KB, this thread will modify the configuration file and regulate the size of data
block back to 4KB. When system recoveries from aging, the original disk queue regulation mechanism
in Helix server works. Figure 36 and Figure 37 demonstrate the bandwidth changes and disk queue
changes before and after rejuvenation. The rejuvenation method is conducted at 3100x15 seconds time
point. From Figure 36, the significant decreasing of disk queue is observed. And after rejuvenation, the
bandwidth increases from around 70kbyte to 100kbyte (See Figure 37). This rejuvenation method is
appropriate and effective when disk concurrent flow is large, namely the request media files are decentralized. \\
\textbf{Rejuvenation based on memory cleaning}

In most real software systems, there are more or less memory cleaning mechanisms (e.g garbage collection
in JVM). \textit{memreap} in Helix server is responsible for cleaning the data blocks which are not referenced
by any request. Parameter \textit{refcount} is used to denote the reference counter. If \textit{refcount} is less than a
threshold (the threshold is 0 for default in Helix server), \textit{memreap} will swap out the data block from
memory. But when file concurrent flow is large, data blocks referenced by these files can't be cleaned in
time. This causes software free radical accumulation and further results in performance degradation. Our
rejuvenation is to enlarge \textit{refcount} when memory exceeds a threshold (e.g \textit{refcount} =15). That means
more data blocks will be swapped out from memory. This experiment is conducted and the bandwidth,
memory and disk changes are observed (see Figure 38-40). At time point 300x15 seconds, rejuvenation
method is carried out. As the workload is light, the memory, bandwidth and disk changes are not so
significant. However we can also see the effectiveness of this rejuvenation method: memory decreases
and bandwidth increases after rejuvenation. It's not good that disk queue increases
slightly after rejuvenation. This side effect happens as the cached data objects are swapped out from memory and new
requests fetch data from disk instead of memory.

From analysis above, the rejuvenation strategies based on strengthening negative feedback loop and
weakening positive feedback loop are effective. And this further proves our software aging mechanism
based on SFR is correct and reasonable. Although these rejuvenation methods can't recover system from
aging state thoroughly, they keep service quality at an acceptable level without service interruption.
\section*{Discussion}
\subsection*{Impact factors of software aging}
Software aging as a complex process is a result of interplay between internal and external
environment of software systems. It may be affected by many factors. Guibert \cite{46} proposed external
pressure $\sum$, internal organization $\Omega $ and temperature $T$ are the impact factors of physical system aging. He used equation: $f = f(\sum,\Omega,T)$ to describe aging. Motivated by his work, we also find out the factors affecting software aging.

The workload from external environment is changing all the time. Literatures \cite{5,47} have proven that
different workloads affect software aging differently. So workload is an impact factor of software aging denoted by the letter $L$. Here the parameters of workload not only include the volume of requests but also the distribution and type of requests. Two workloads are different if even one parameter is different. Formally, let $({w_{1}}^{i}, {w_{2}}^{i}, \cdots , {w_{n}}^{i})$ denote the workload of $L_{i}$, and  $({w_{1}}^{j}, {w_{2}}^{j}, \cdots , {w_{n}}^{j})$denote the workload
of $L_{j}$. If $\exists m, m\in (1,2,\cdots ,n), w_{m}^{i}\neq w_{m}^{j})$ then $L_{i} \neq L_{j}$. Intuitively, software aging is affected by the organization of software system just as different organisms have different aging speeds. By the word organization, we mean the functional and non-functional components consisting the software system. If a software system has a perfect fault-tolerant or adaptive adjusting mechanism, the software free radicals may be cleaned in time and software aging could be mitigated significantly. Otherwise software aging could be worsen. So the organization of the software system is also a factor of software aging denoted by $O$. Another important factor is computing resource including physical resource (i.e. cpu cycles, memory or disk space) and logical resource (i.e. files hands or locks). If a software
system doesn't have enough resources, the performance could decrease. Here we use $R$ to represent this factor. Then the software aging process could be defined as function of $L$, $O$ and $R$: $f = f(L,O,R)$. During software running, $L$, $O$ and $R$ change with time. For example workload changes from low to high, adaptive adjusting mechanism changes from \textit{on} to \textit{off} and resource changes from \textit{much} to \text{little}. Therefore software aging is a dynamic process rather than a static process: 
$$f = f(L(t),O(t),R(t))$$
In order to study how a factor affects software aging, two other factors should be fixed. Then find the relationship between the studied factor and software aging through changing this factor. For example, if we want to study the relationship between workload and software, we should fix software system and system resources, and then change workload. But if we want to study how software organization affects software aging, we should choose different software systems besides fixing workload and resources. Actually in real software systems, it's hard to do these experiments as workload, organization and resources are interleaved mutually. First, workload changes can cause resource changes. More workloads need more resources. Second, workload changes can trigger some adaptive adjusting mechanisms such as load balancing. And adjusting mechanisms changes also causes workload changes. Third, resources changes can also trigger some adaptive adjusting mechanisms such as garbage collection in JVM. Meanwhile adaptive adjusting mechanism changes may result in resources changes. The relationship among these three factors is demonstrated in Figure 41. To a specific software system such as Helix server implemented in our paper, total system resources and organization are fixed, what we can control is the workload. So in this paper, we only discussed how workload affects software aging. The different characteristics among variant software systems and how resources affect software aging will be discussed further in our future work.

The completeness and orthogonality of aging impact factors are not discussed in this paper. There may be some other factors, but as far as we know, the aforementioned factors have covered nearly all the known aging impact factors. For example, aging-related bug \cite{17,18,19} could be attributed to the organization factor due to the defects in the source code and memory leak \cite{20} could be attributed to the resource factor. At last, we  only propose a new perspective to explain the mechanism of software aging `stealing' ideas from biological studies rather than overthrow the previous studies about software aging.
\section*{Conclusion}
Software aging has been a mysterious problem to software researchers and developers. Although this
phenomenon has been studied for nearly two decades, the fundamental mechanism of software aging is still not discovered. Through a interdisciplinary study, this paper finds the common points between software aging and biological aging. Motivated by these common points we propose a new perspective to explain the mechanism of software aging built on free radical aging theory which is widely accepted in biological studies.The concept of `software free radical' is presented and a new conclusion that the software aging process is somehow attributed to the accumulation of software free radical is stated. Based on the metabolic process of software free radical, a relatively general aging model which combines negative feedback loop and positive feedback loop is developed. Moreover the proposed mechanism and aging model are validated in several well known open source software systems. Once the mechanism is discovered the cost-effective software rejuvenation will be conducted. In this paper, several simple but effective software rejuvenation strategies are implemented and the experimental results further show the effectiveness of our method. Although this work is still in infancy, it is a first try to combine software aging and biological aging. And lots of unresolved problems need to be discussed deeply. At last, we hope this paper a seminal work to motivate more valuable researches in the interdisciplinary field of computing science and biological science.
\section*{Materials and Methods}
\subsection*{Model of software aging}
In this part a quantitative description of software aging is given. From analysis above, the metabolic
process of software free radical is clear. But it's hard to measure the accumulation speed of software free
radical directly as it is information, an abstract concept not a real entity. It's well known that computing
resources (i.e. memory, disk) are the carriers of information. Therefore we indirectly measure software
free radical accumulation speed through measuring resources consumed by software free radical. Using
$Resource_{SFR}$ to represent resources taken by software free radical, the metabolic process of software free
radical could be substituted by $Resource_{SFR}$ metabolic process. This process is shown in Figure 42, the
upper part is the positive feedback loop and the lower part is the negative feedback loop.
Software free radical cleaning mechanisms including garbage collection and other adaptive adjusting
mechanism make up the negative feedback loop. They mitigate software aging through cleaning software
free radicals and releasing resources taken by these free radicals. Following \cite{46} , one order negative feedback
function is adopted to describe this process:
\begin{equation}
y(t)=K(1-e^{-bt})
\end{equation}
Here $y(t)$ represents the degree of aging at different time, $K$ is a constant, $b$ denotes the ability to mitigate
software aging. In most software systems, $b$ is not a constant but a decreasing function of time. In the
initial running stage, software adaptive adjusting mechanism could clean software free radicals in time.
The system has a strong power to keep normal running. But with software free radical accumulating,
the system can't clean them in time and become weak to keep normal running. Here $b$ is simply defined
as the summary of inverse of time t and a constant $v$. That is:
\begin{equation}
b(t)=\frac{1}{t}+v
\end{equation}
In the positive feedback loop, software free radical increases with time. One order positive feedback is
implemented to describe this process: 
\begin{equation}
z(t)=z_{0}e^{at}
\end{equation}
Here $z(t)$ represents the degree of aging at different time, $z_{0}$ means initial aging level, a denotes the
ability to speed up software aging, it's a constant. As mentioned above, software aging integrates the
negative feedback loop and positive feedback loop. So we integrate equation (5),(6)and equation (7) in
the multiply form and get the software aging model as follows:
\begin{equation}
f(t)=z_{0}e^{at}(1-e^{-(1/t+v)t})
\end{equation}
In the integrated feedback loop process, we make an assumption that there are only one negative feedback loop and only
one positive feedback loop. In reality, software system may have multiple negative or positive feedback
loops. In this situation, a more complex aging model is given:
\begin{equation}
f(t)=y_{1}(t)z_{1}(t)+y_{2}(t)z_{2}(t)+,\cdots ,y_{n}(t)z_{n}(t))
\end{equation}
Here $y_{i}(t)$ represents different negative feedback loops and $z_{i}(t)$ represents different positive feedback
loops. Apparently it's extraordinarily hard to estimate the model parameters of equation (9). Therefore,
this paper employs a more generic and simple differential equation to describe software aging:
\begin{equation}
\frac{\dot{f}(t)}{f(t)}=\alpha + \frac{\beta }{t}
\end{equation}
$\alpha $ denotes the aging speed generated by positive feedback loop, it is a constant. $\frac{\beta }{t}$ is the aging speed affected by negative feedback loop, it's a monotonously decreasing function of time, $\beta $ is a constant. $\dot{f}(t)$, the derivative of $f(t)$, is the total software aging speed. When $\beta =0$, namely there is no negative feedback loop, $\dot{f}(t)=\alpha (t)$ which means software aging speed is proportional to current software aging degree; when $\alpha = 0$ namely there is no positive feedback loop, $\dot{f}(t)=\beta (f(t)/t)$ which means software
aging speed is decreasing with time. If $t\rightarrow \propto, \dot{f}(t)=0$, so without positive feedback loop software aging phenomenon will disappear through self adjusting given enough time. Resolve equation (10), we get the explicit expression of software aging:
\begin{equation}
f(t)=Ke^{\alpha t}t^{\beta}
\end{equation}
\section*{Experiment Methodology}
We conducted many full-fledged experiments on three well known open source platforms: TPC-W benchmark, apache HTTP web server and Helix stream media server.

TPC-W \cite{48} is a transaction processing benchmark. The experiment is performed in a controlled environment mimicing the real activities of book shopping. The
performance metric reported by TPC-W is the number of web interactions processed per second. Multiple
web interactions are used to mimic the activities of a book retail store, and each interaction is subject to a
response time constraint. As a standard multi-tiered application, TPC-W has a web presentation tier,
an application processing tier and a database tier. In this paper, we deployed the TPC-W benchmark in
two dedicated physical machines: a Tomcat application server which is used to process the web requests
and servlet applications and a MySQL database server which is used to store the data. To accelerate
software aging, we leverage the browser emulator generation program to emulate a higher concurrent clients, say 300 clients and set the think time (the interval between
two consecutive requests) a smaller value, say 0.5s. After several hours running, the emulators stopped
working. After investigating the two running servers, we found the memory utilization of MySQL
server exceeded the whole physical main memory and finally crashed. We have repeated the whole process
several times and observed the same results. 

Apache web server \cite{49} is an open source and widely used web server which is used to transfer web
pages through HTTP protocol. It has been reported that the response time of web pages increased after
a long time running \cite{7,14}. We deployed apache web server on a dedicated server machine and used five
client machines to issue the requests. In order to emulate the real situation, we collected more than
40 thousands of web pages from www.ask.com  site and leveraged Siege \cite{50}, a web site stressing tool, to
generate requests randomly. The response time of each web page is logged during the tests.

Helix server \cite{51} as a mainstream streaming media software system transmits video and audio between
client and server using RTSP/ RTP protocol. It is consisted of many components including scheduling component, monitoring component, memory adjusting component (memreaper), media file processing components and other components. From the scale and complexity, Helix server could be regarded as a complex system. At present, there is no standard streaming media benchmark. So this paper develops a client emulator named \textit{`HelixClientemulator'} employing RTSP and RTP protocols.
\textit{`HelixClientemulator'} contains three threads, one for audio processing, one for video processing and
the other for session management. It can generate multiple clients to request media files on Helix server.

100 rmvb media files with different rates are deployed on the Helix server. In order to study how
workload affects software aging, this paper leverages \textit{`HelixClientemulator'}  to generate different workloads. The workload
is identified by a tuple containing six elements: \textit{(client count, file dist, file object, file max object, sleep time, file difference)}. The
detailed description of these attributes is stated in Table 2. In every experiment, client count is fixed. If a
client finishes file transmission , a new client will be created. In order to avoid the side effect from previous
experiment, Helix server is restarted before every experiment.

To make sure the volume of workload is within the capacity of software system, capacity test is performed.
When the request number arrives 900, Helix server is restarted automatically and the transmission speed
drops immediately shown in Figure 43. So 900 is the limit capacity of the test, the following experiments
will not exceed this limit. The bandwidth of every simulated client doesn't exceed 140kbps, so the
total bandwidth is no more than 126Mbps within the service capacity of our exchanger. Therefore
the transmission ability is not a bottleneck in our test bed. During the software system running,
thousands of performance metrics and events are generated.The main task of Helix server is to deliver
multimedia files. For this kind of software, file cache module is a very important component to optimize
the transmission speed. This module has significant impact on memory and disk queue length. So two
performance metrics \textit{AvgDiskQueueLength} and \textit{SystemCacheResidentBytes} should be monitored. In
Helix server, SFR mainly takes up memory resource. Hence the metric \textit{ProcessWorkingSet} which
indicates memory exhaustion of Helix server process is used to identify  the accumulative process of
SFR. And the transmission speed is a critical QoS (Quality of Service) indicator. Therefore the metric
\textit{AveragebandwidthOutputPerPlayer(kbytes)} is used to show whether software aging happens.
\section*{Acknowledgments}
This paper is a primary interdisciplinary trial. During this long and tough research, we got great support
from processor Jiankang liu and jiangang long who are both the faculties of School of Life Science \&
Technology of Xi'an jiaotong university. And our work is supported by the Fund: Research on Networked
Software Aging Mode and Rejuvenation Approach (Grant No. 60933003) which is sponsored by the Key
Project of National Natural Science Foundation of China.
\bibliography{reference}
\section*{Figure Legends}
\begin{figure}[!ht]
\begin{center}
\includegraphics[width=6in]{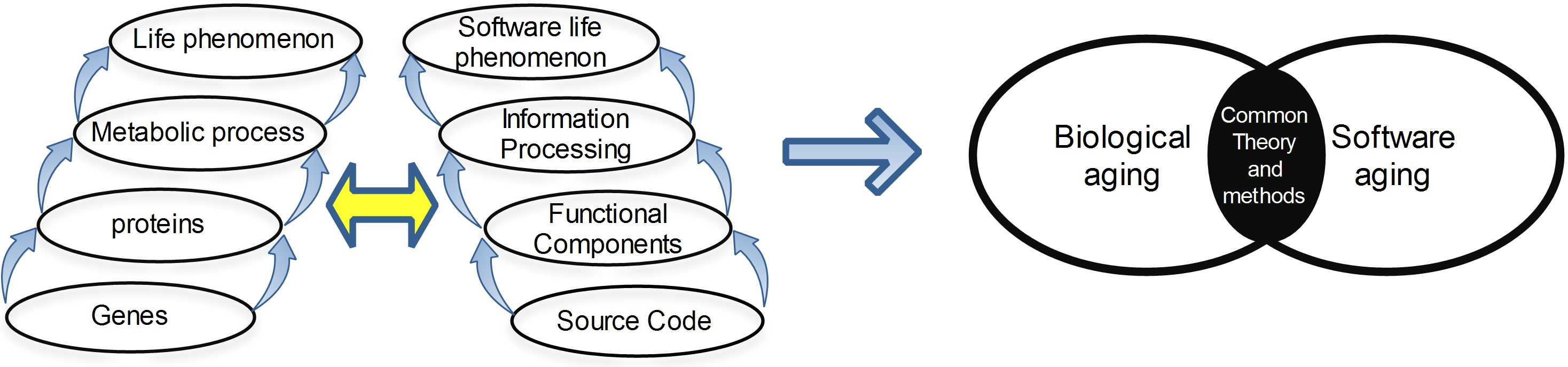}
\end{center}
\caption{
{\bf The equivalent concept stack between organism and software. }  This figure demonstrates that many similarities exist between these two entities at different concept granularity. A
reasonable conjecture could be concluded: there may be common theory and methods between software
aging and biological aging. The interdisciplinary studies may give us some insights to software aging.
}
\label{2}
\end{figure}

\begin{figure}[!ht]
\begin{center}
\includegraphics[width=4in]{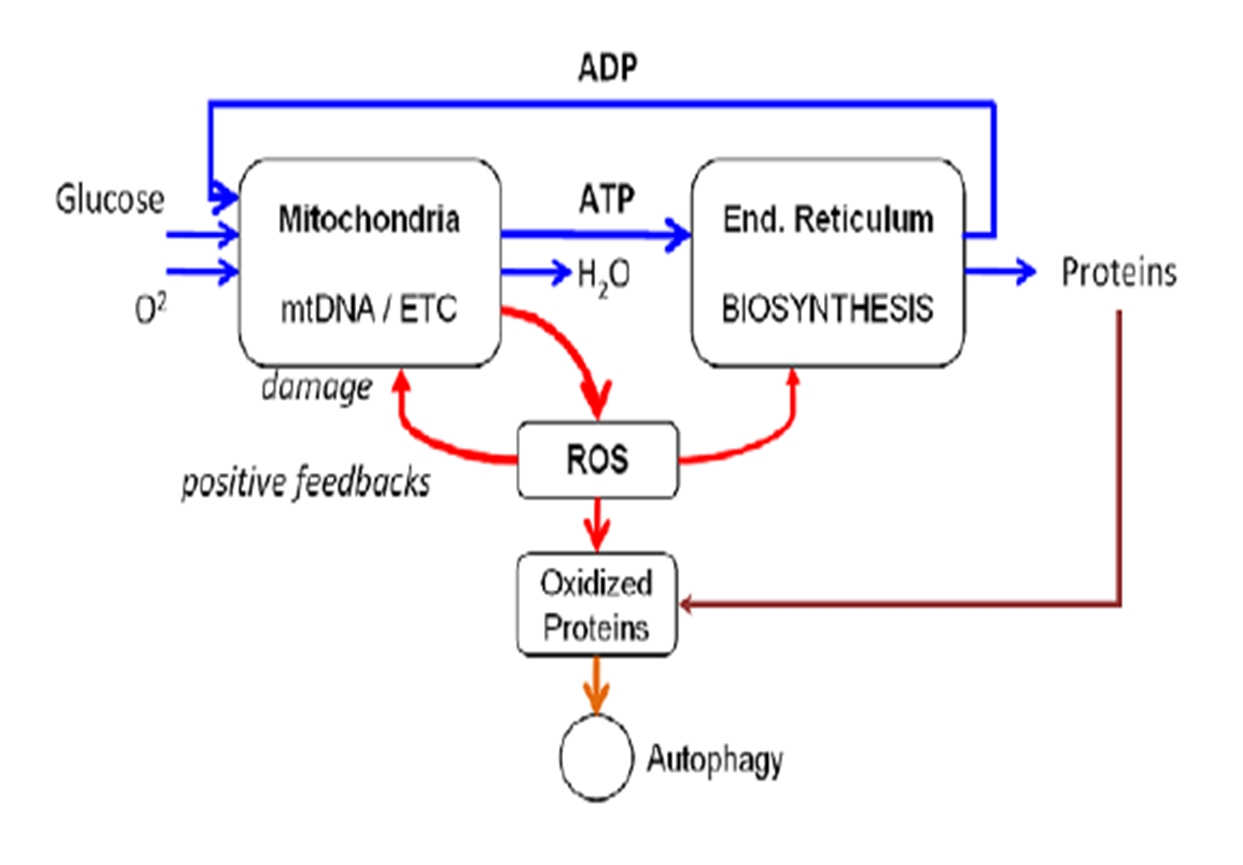}
\end{center}
\caption{
{\bf Graph of a positive feedback-loop motif. }  Cited from literature \cite{44}.
}
\label{free_radical}
\end{figure}

\begin{figure}[!ht]
\begin{center}
\includegraphics[width=4in]{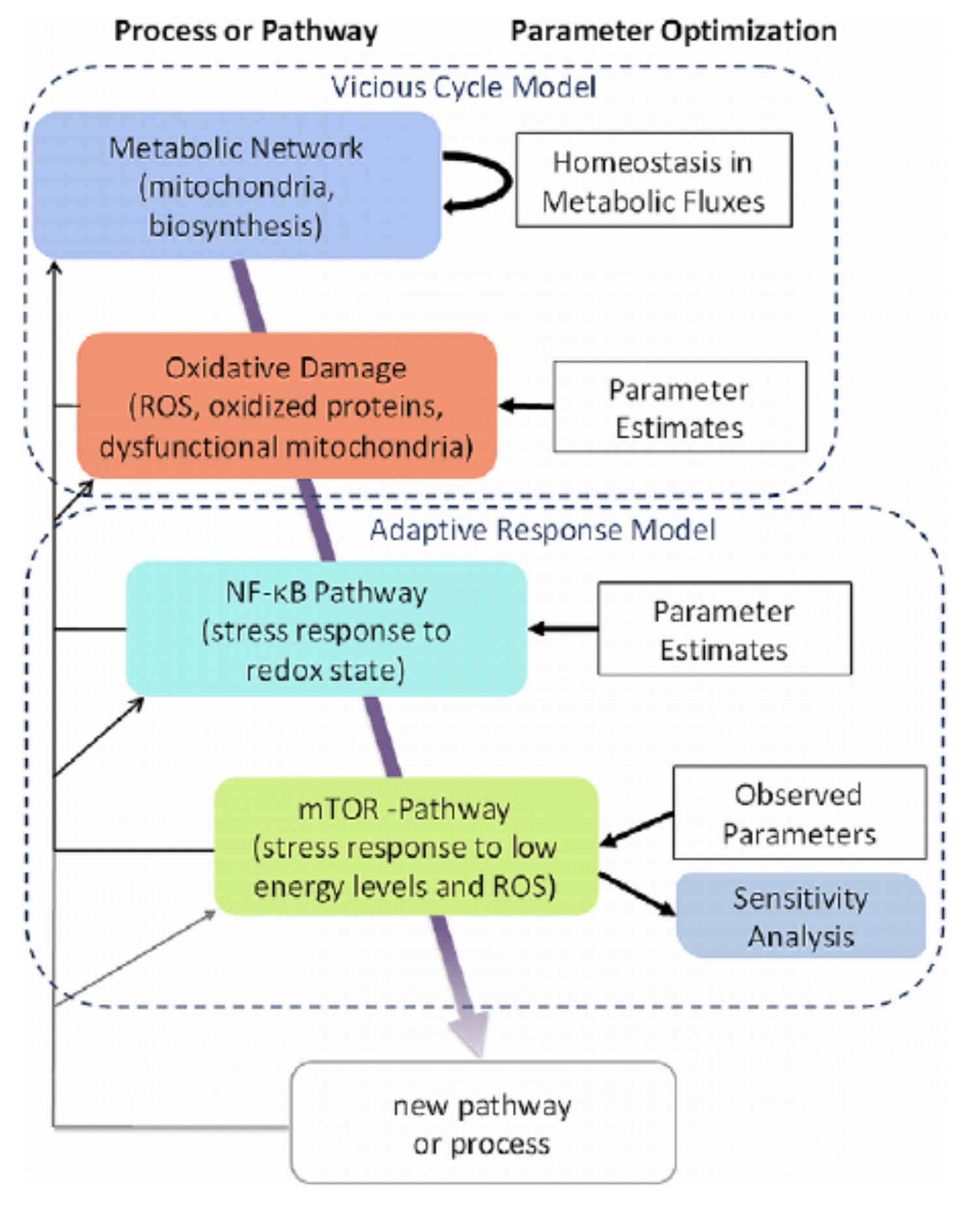}
\end{center}
\caption{
{\bf Overview of the steps taken to assemble a generic cell aging model. }  Cited from literature \cite{44}.
}
\label{free_radical}
\end{figure}

\begin{figure}[!ht]
\begin{center}
\includegraphics[width=4in]{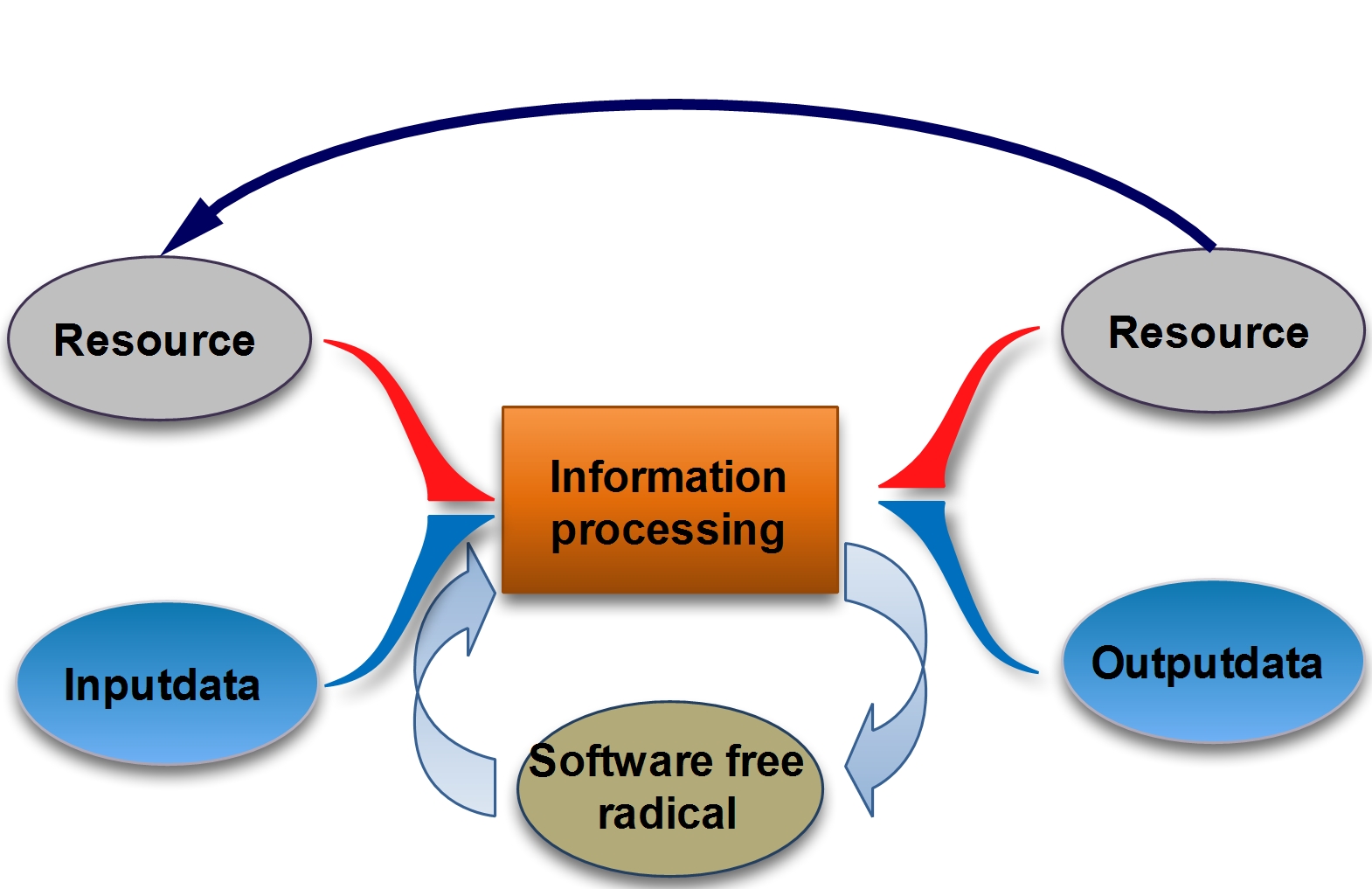}
\end{center}
\caption{
{\bf The procedure of `software free radical' metabolic in running software systems} Software systems utilize computing resources such as CPU, memory to transform the input data (e.g. HTTP requests) to the specific output data (e.g. web content) and simultaneously recall the computing resources when the transformation is done. During this procedure, SFR is also generated.
}
\label{free_radical}
\end{figure}

\begin{figure}[!ht]
\begin{center}
\includegraphics[width=4in]{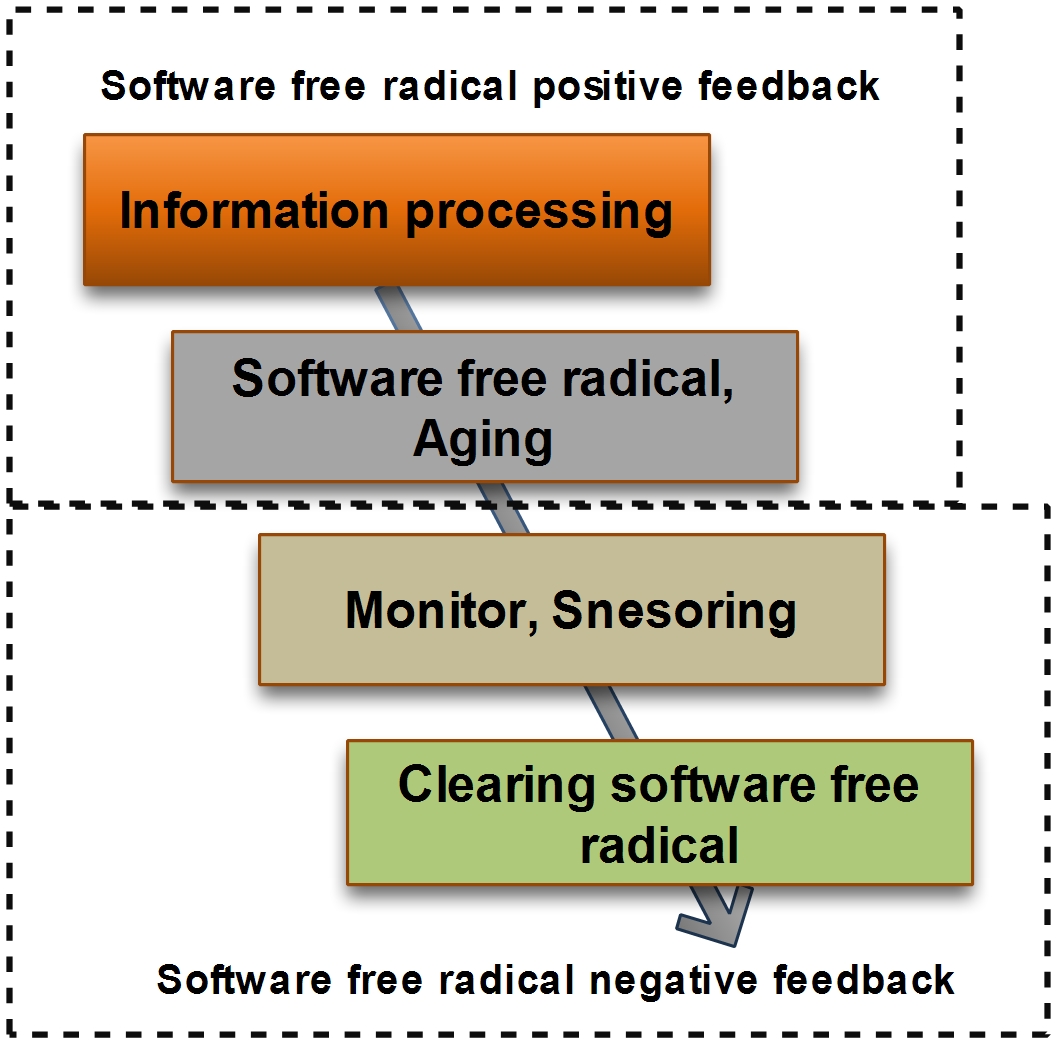}
\end{center}
\caption{
{\bf The negative and positive feedback loops of `software free radical' in running software systems}.
}
\label{free_radical}
\end{figure}

\begin{figure}[!ht]
\begin{center}
\includegraphics[width=4in]{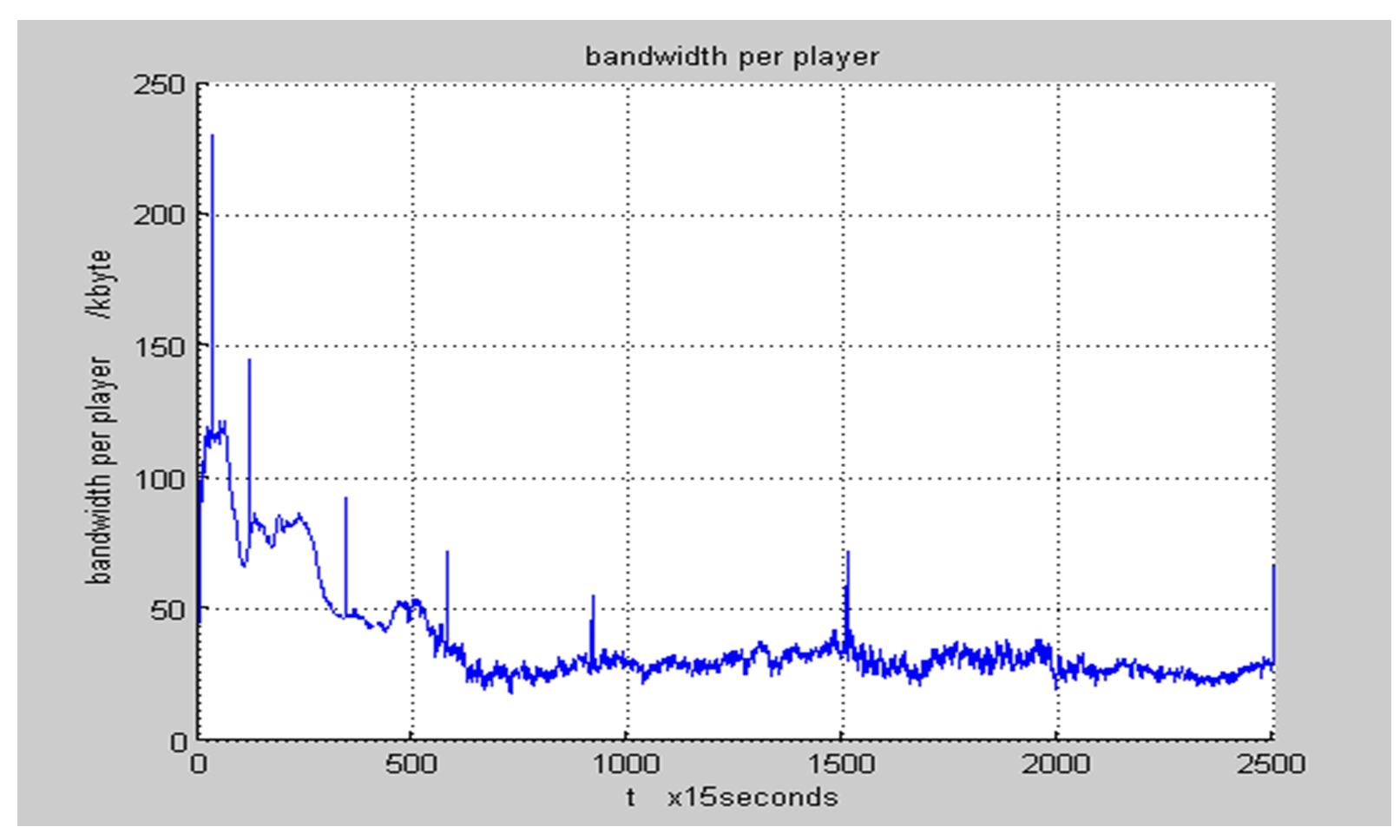}
\end{center}
\caption{
{\bf The degradation of `bandwidth per player' metric of Helix server under workload $w_{1}$} .
}
\label{free_radical}
\end{figure}

\begin{figure}[!ht]
\begin{center}
\includegraphics[width=4in]{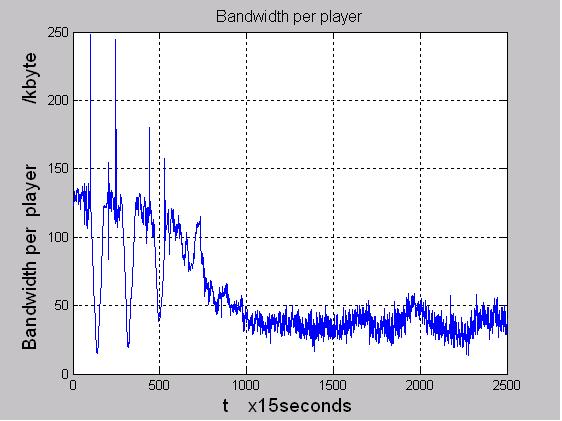}
\end{center}
\caption{
{\bf the degradation of `bandwidth per player' metric of Helix server under workload $w_{2}$}.
}
\label{free_radical}
\end{figure}

\begin{figure}[!ht]
\begin{center}
\includegraphics[width=4in]{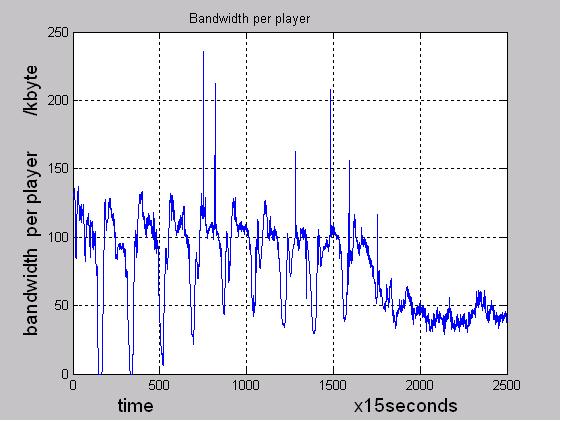}
\end{center}
\caption{
{\bf The degradation of `bandwidth per player' metric of Helix server under workload $w_{3}$}.
}
\label{free_radical}
\end{figure}

\begin{figure}[!ht]
\begin{center}
\includegraphics[width=4in]{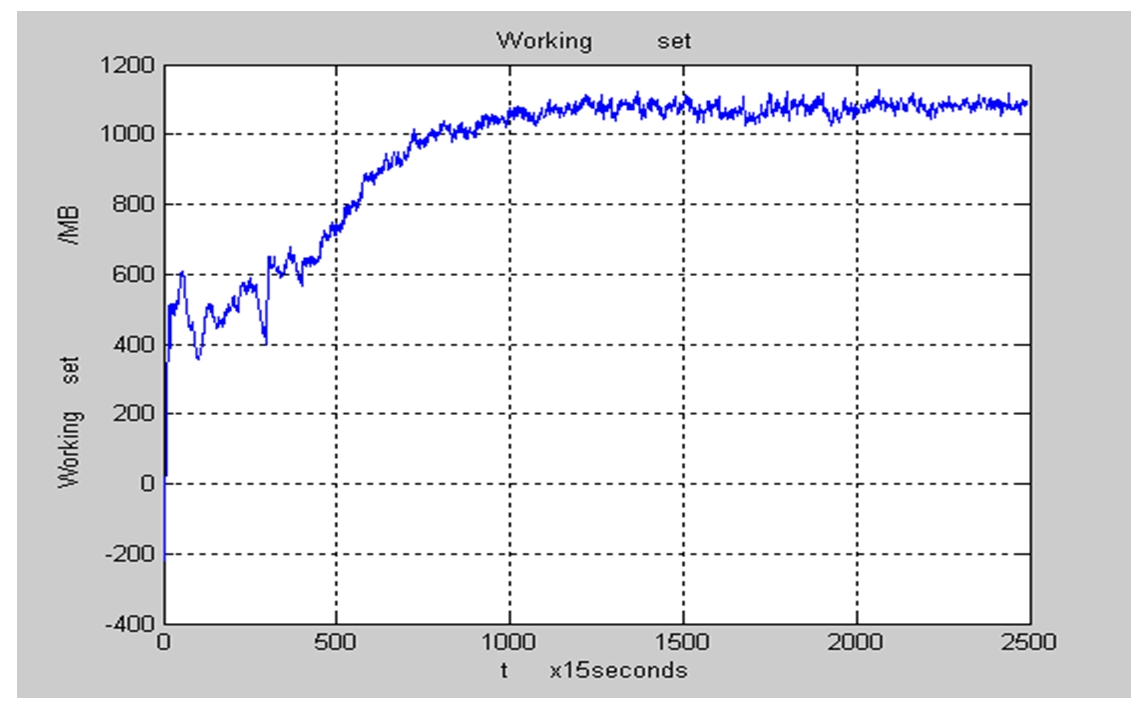}
\end{center}
\caption{
{\bf The increasing of `working set' metric of Helix server under workload $w_{1}$ }.
}
\label{free_radical}
\end{figure}

\begin{figure}[!ht]
\begin{center}
\includegraphics[width=4in]{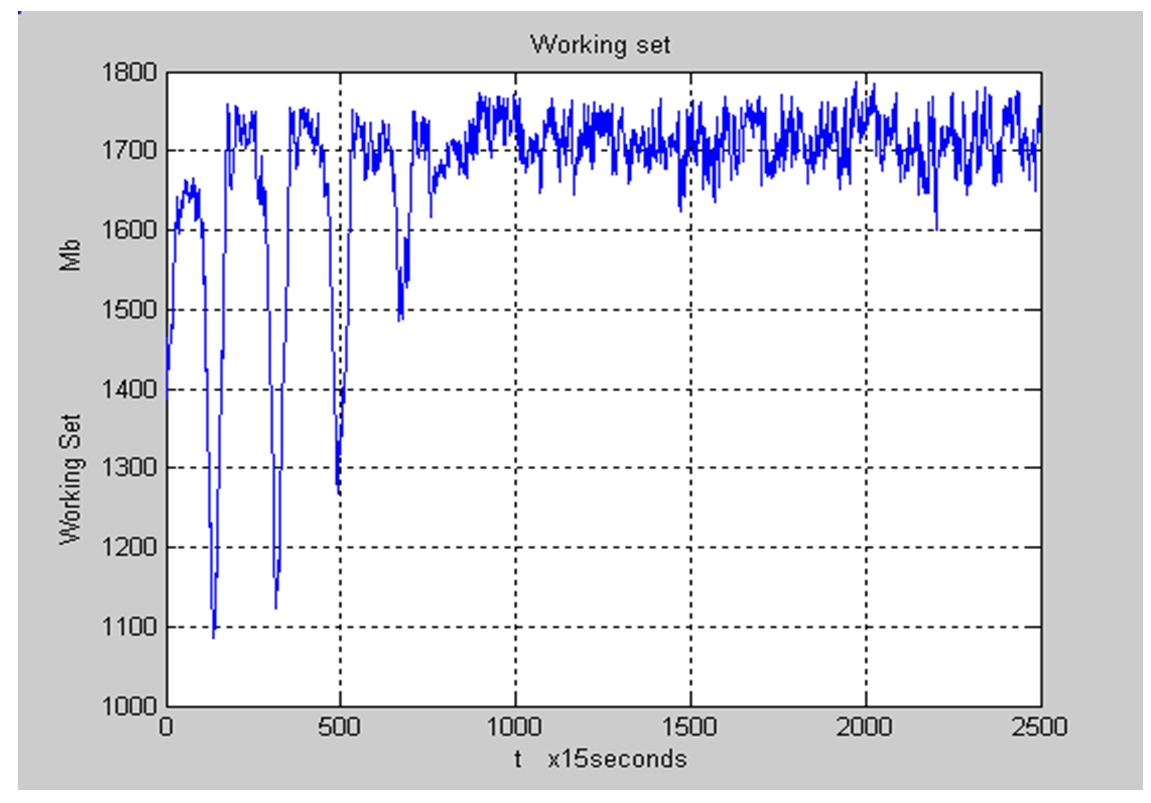}
\end{center}
\caption{
{\bf The increasing of `working set' metric of Helix server under workload $w_{2}$ }
}
\label{free_radical}
\end{figure}

\begin{figure}[!ht]
\begin{center}
\includegraphics[width=4in]{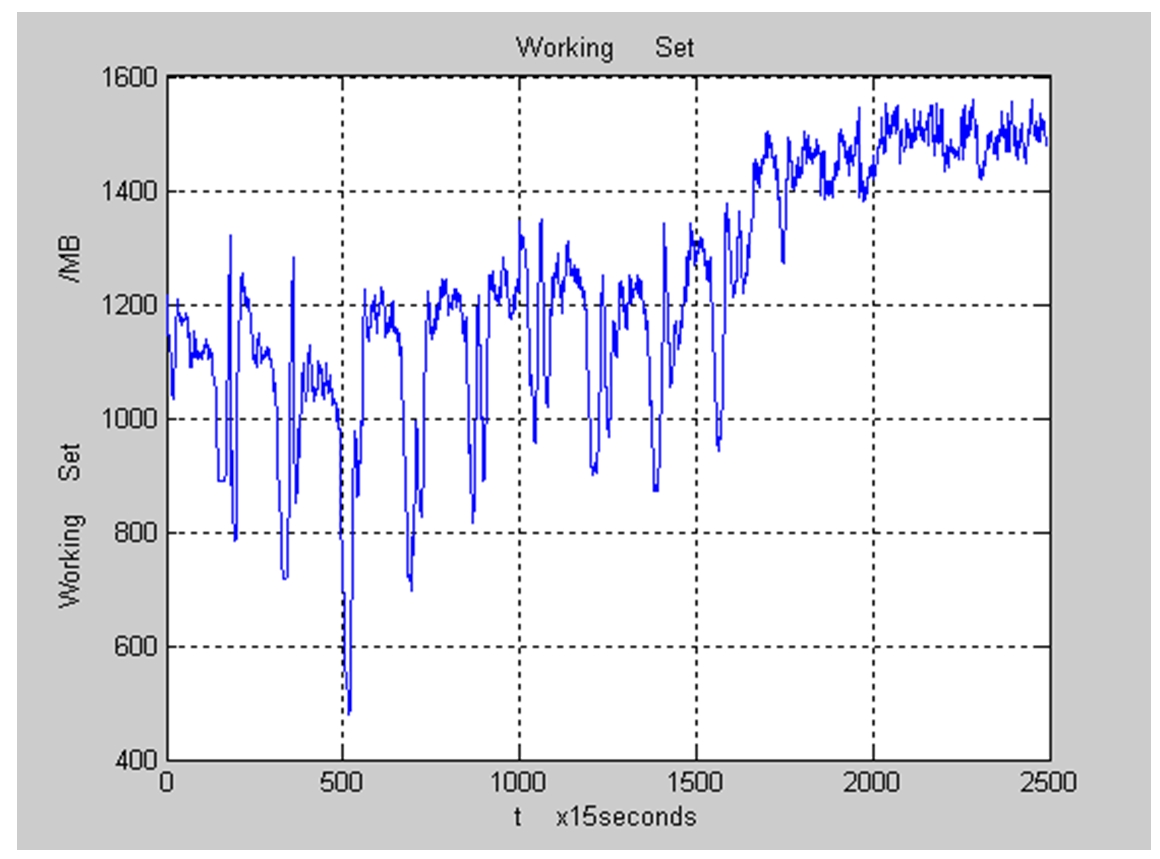}
\end{center}
\caption{
{\bf The increasing of `working set' metric of Helix server under workload $w_{3}$ }.
}
\label{free_radical}
\end{figure}

\begin{figure}[!ht]
\begin{center}
\includegraphics[width=4in]{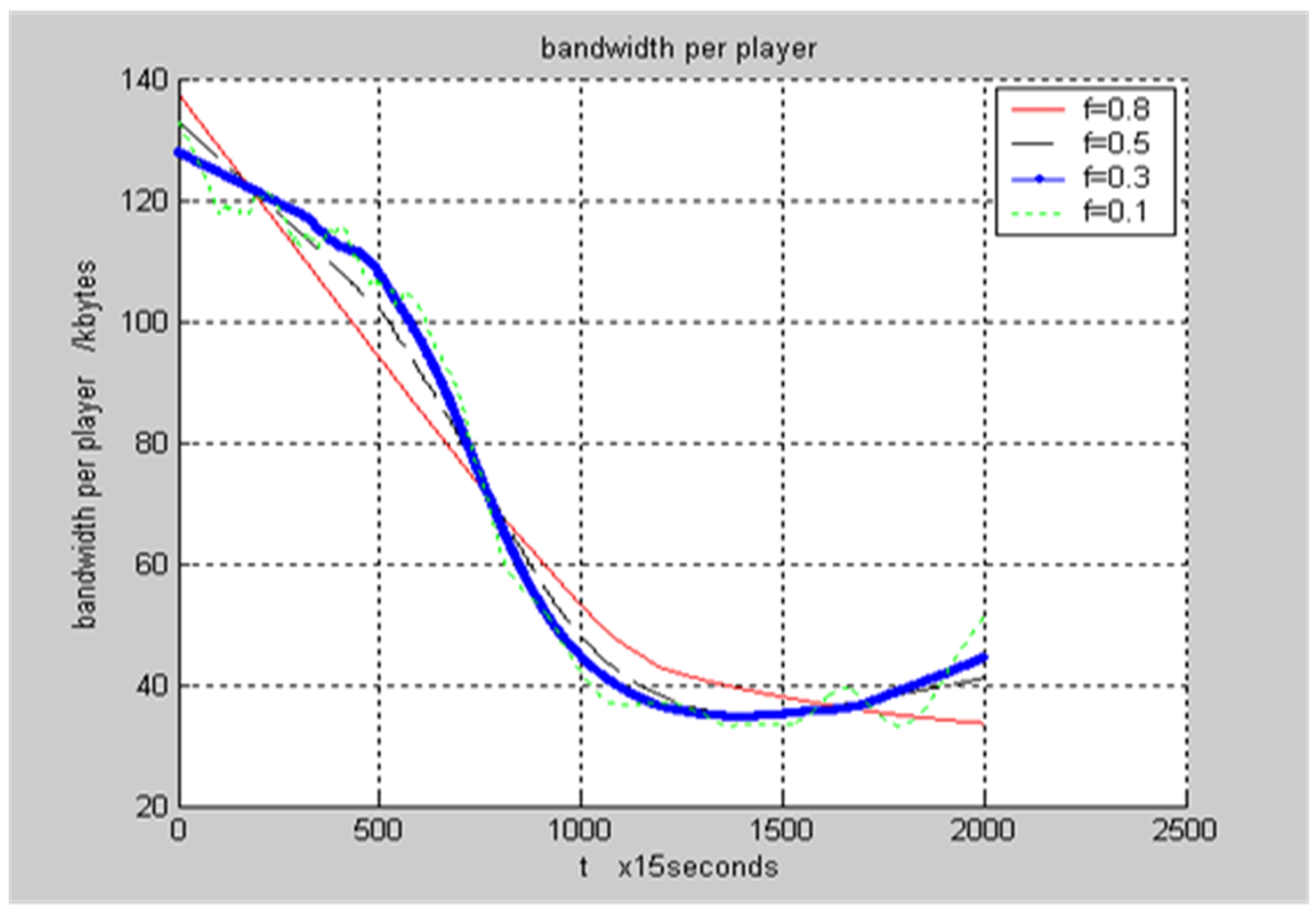}
\end{center}
\caption{
{\bf The aging trend of `bandwidth per player' of workload $w_{2}$ smoothed by \textit{Lowess} under different $f$ value in Helix server benchmark.}
}
\label{free_radical}
\end{figure}

\begin{figure}[!ht]
\begin{center}
\includegraphics[width=4in]{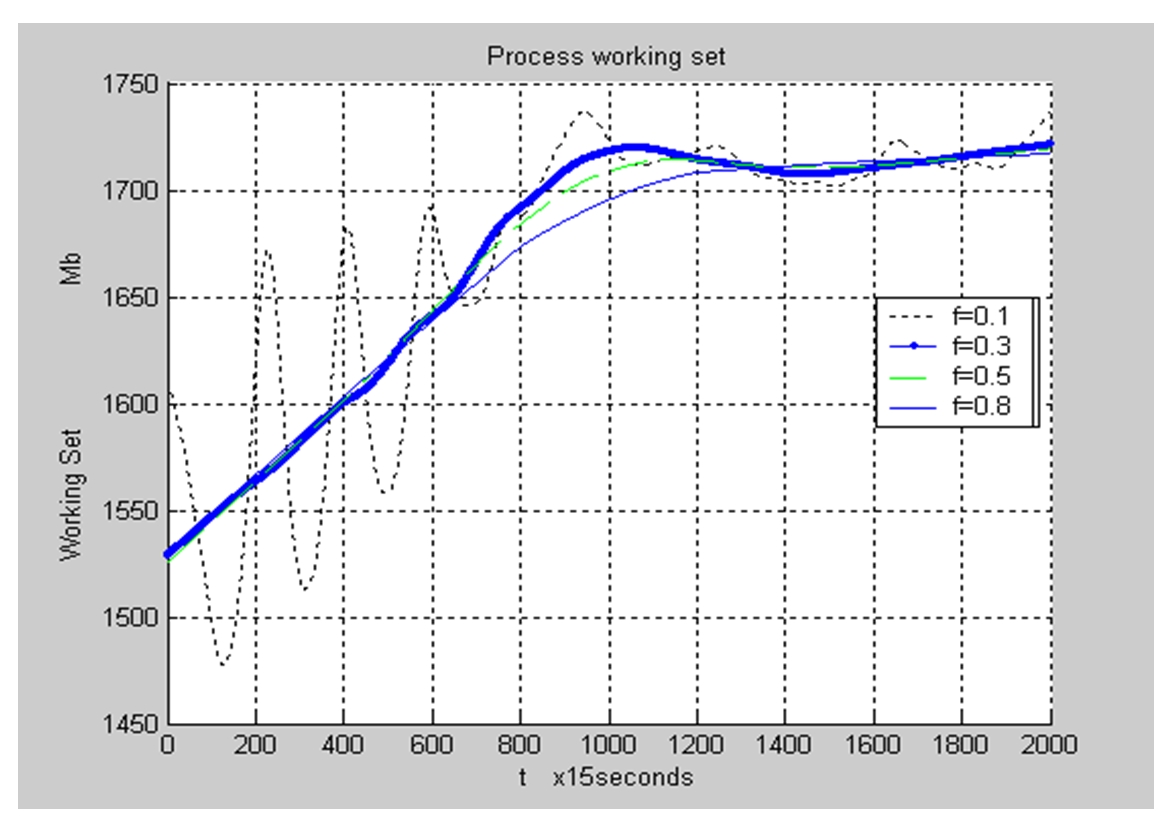}
\end{center}
\caption{
{\bf The aging trend of `working set' of workload $w_{2}$ smoothed by \textit{Lowess} under different $f$ value in Helix server benchmark.}.
}
\label{free_radical}
\end{figure}

\begin{figure}[!ht]
\begin{center}
\includegraphics[width=4in]{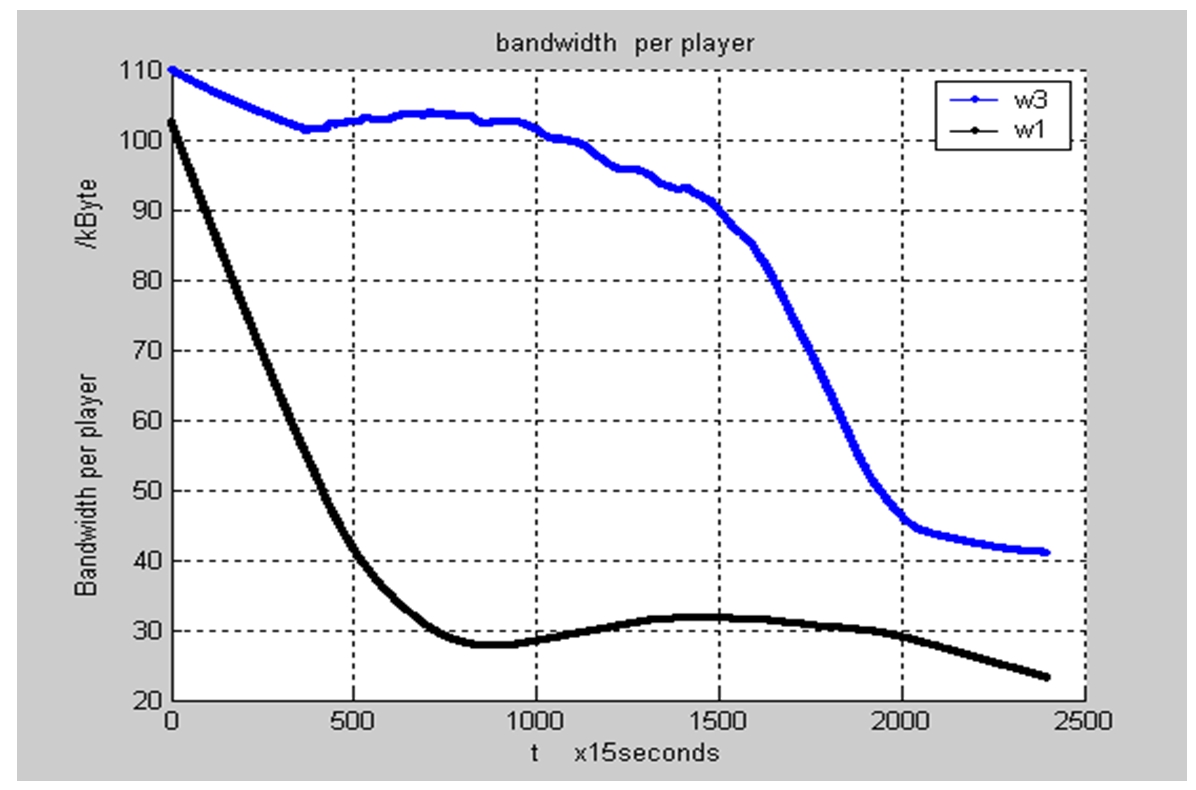}
\end{center}
\caption{
{\bf The aging trend of `bandwidth per player' of workload $w_{1}$  and   $w_{3}$ smoothed by \textit{Lowess} when $f=0.3$ in Helix server benchmark. }.
}
\label{free_radical}
\end{figure}

\begin{figure}[!ht]
\begin{center}
\includegraphics[width=4in]{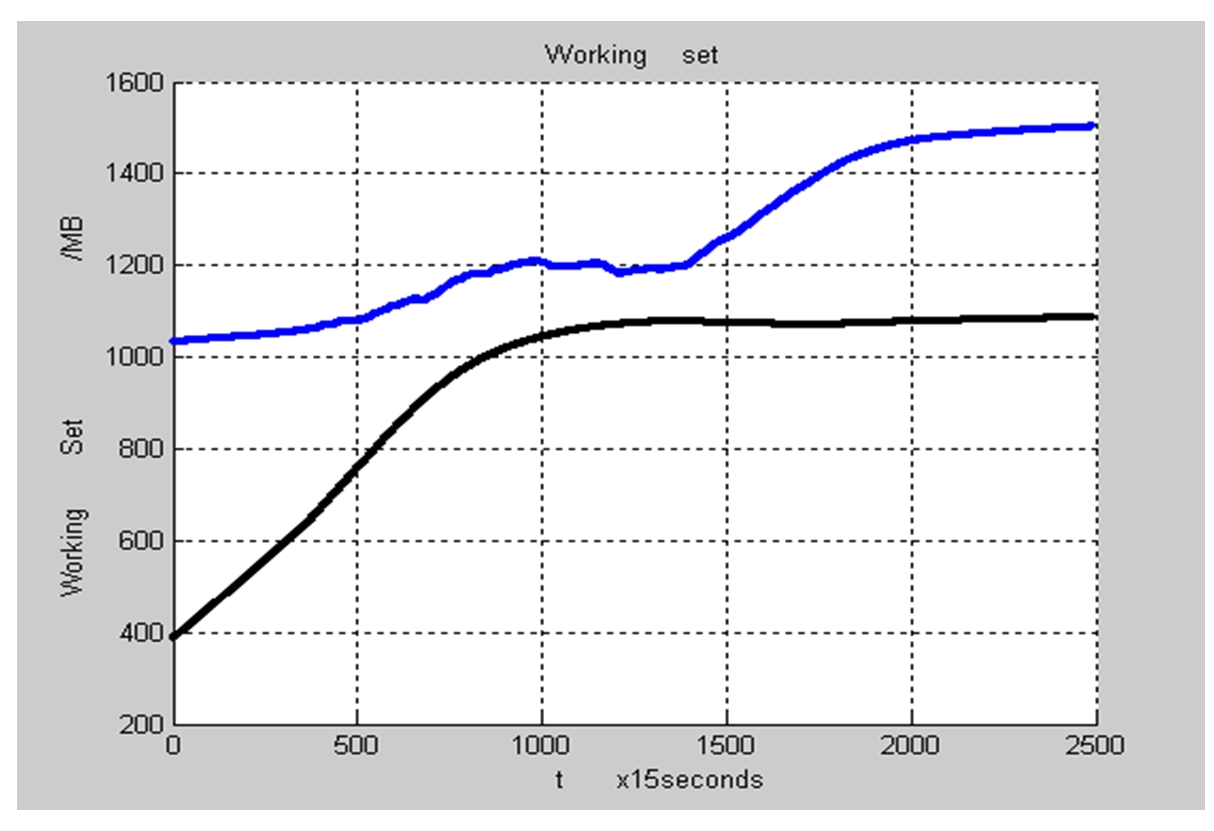}
\end{center}
\caption{
{\bf The aging trend of `working set' of workload $w_{1}$  and  $w_{3}$ smoothed by \textit{Lowess} when $f=0.3$ in Helix server benchmark.  }
}
\label{free_radical}
\end{figure}

\begin{figure}[!ht]
\begin{center}
\includegraphics[width=4in]{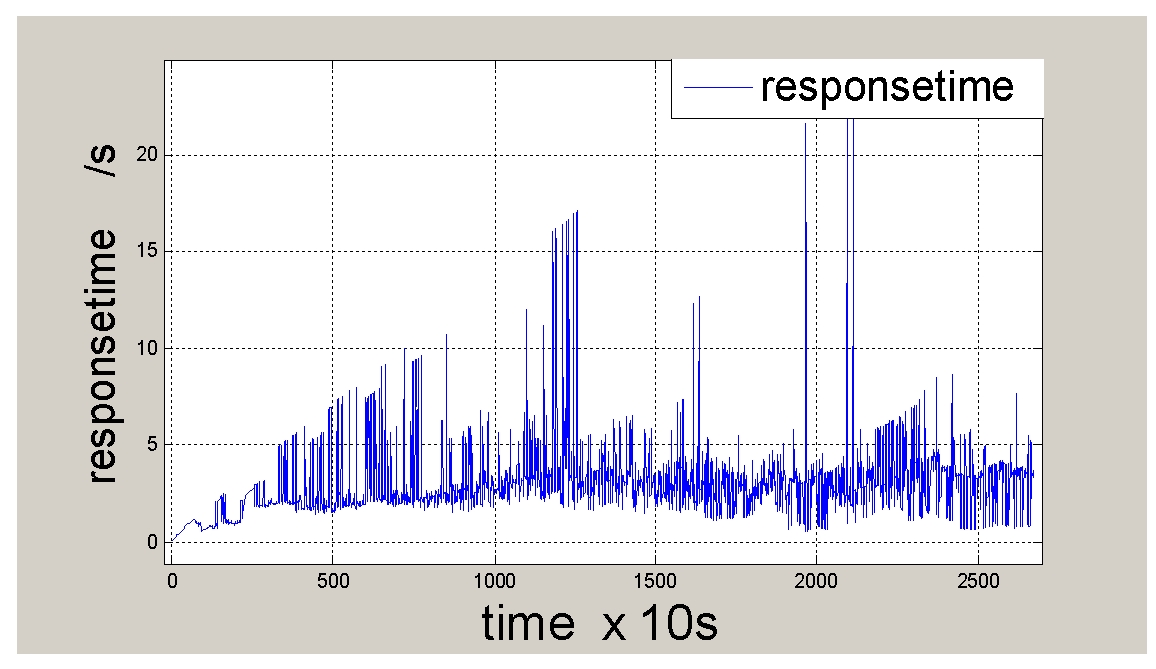}
\end{center}
\caption{
{\bf The increasing of `response time' metric of web server}.
}
\label{free_radical}
\end{figure}

\begin{figure}[!ht]
\begin{center}
\includegraphics[width=4in]{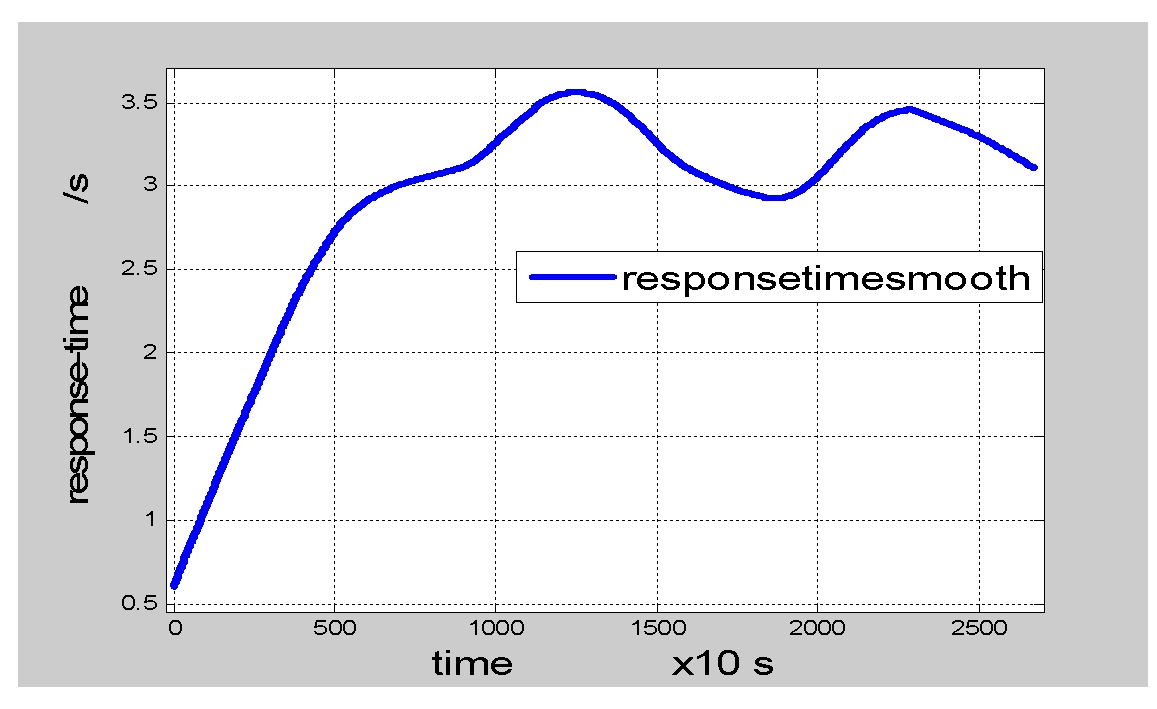}
\end{center}
\caption{
{\bf The aging trend of `response time' metric smoothed by \textit{Lowess} when $f=0.3$ in web server benchmark}.
}
\label{free_radical}
\end{figure}

\begin{figure}[!ht]
\begin{center}
\includegraphics[width=4in]{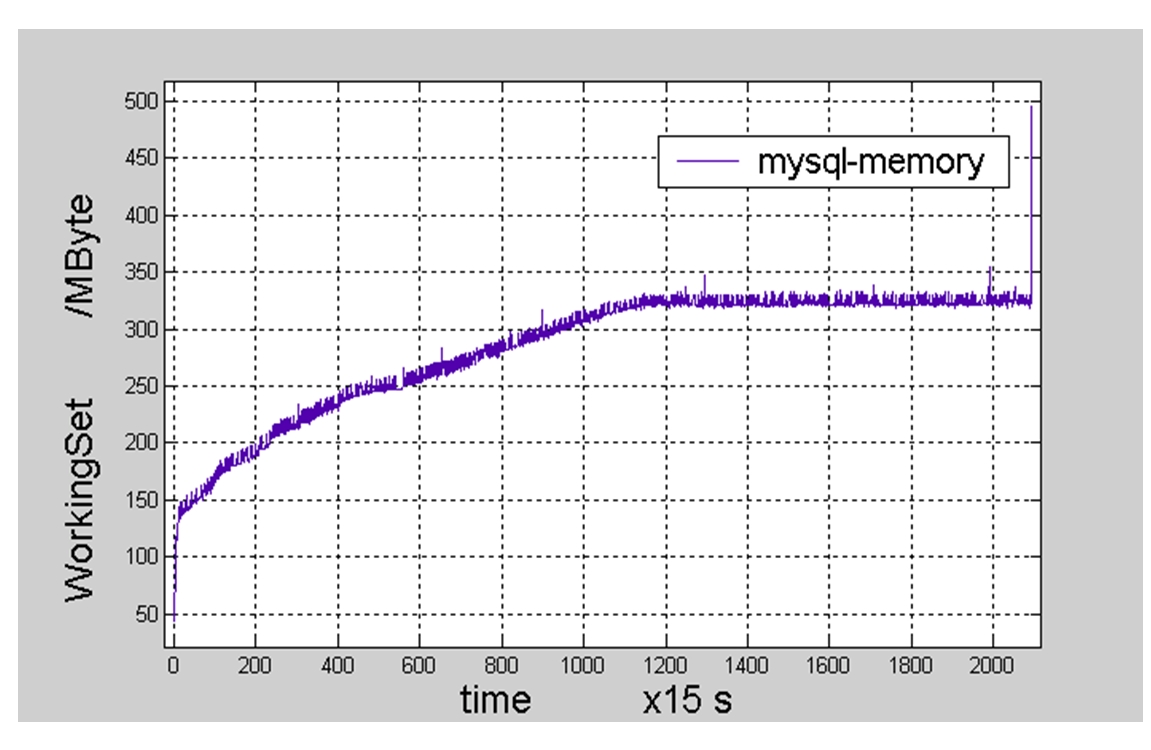}
\end{center}
\caption{
{\bf  The increasing of `working set' metric of mysql server}
}
\label{free_radical}
\end{figure}
\clearpage
\begin{figure}[!ht]
\begin{center}
\includegraphics[width=4in]{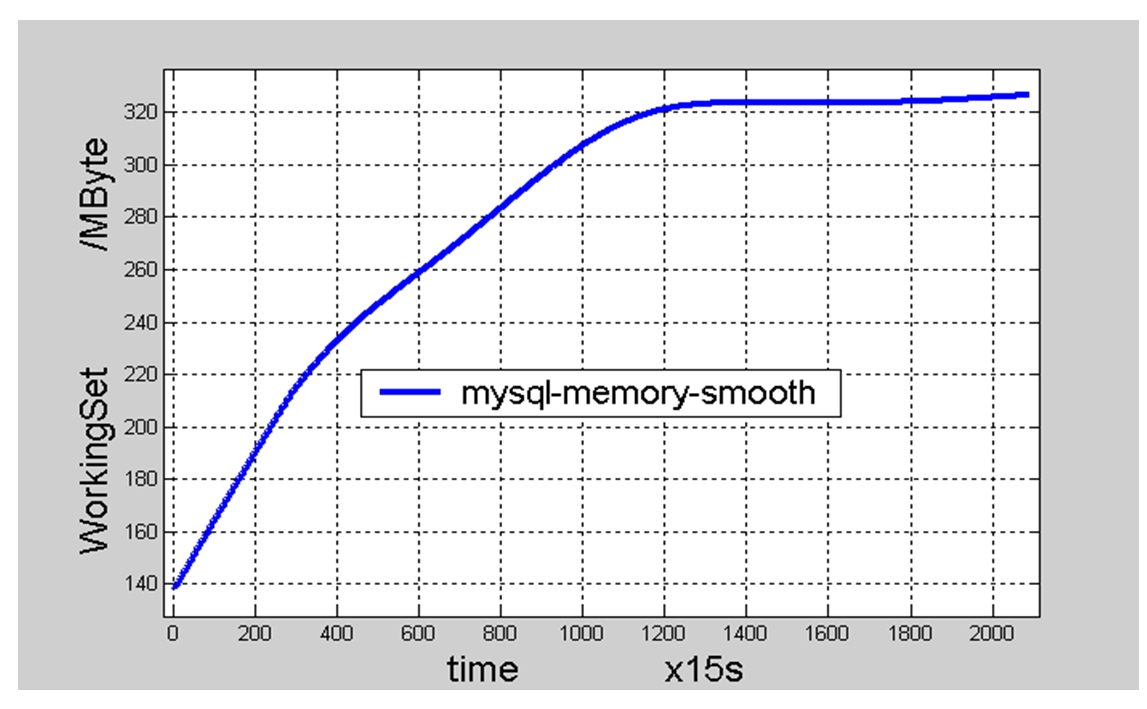}
\end{center}
\caption{
{\bf The aging trend of `working set' metric smoothed by \textit{Lowess} when $f=0.3$ in TPC-W benchmark.} 
}
\label{free_radical}
\end{figure}
\begin{figure}[!ht]
\begin{center}
\includegraphics[width=4in]{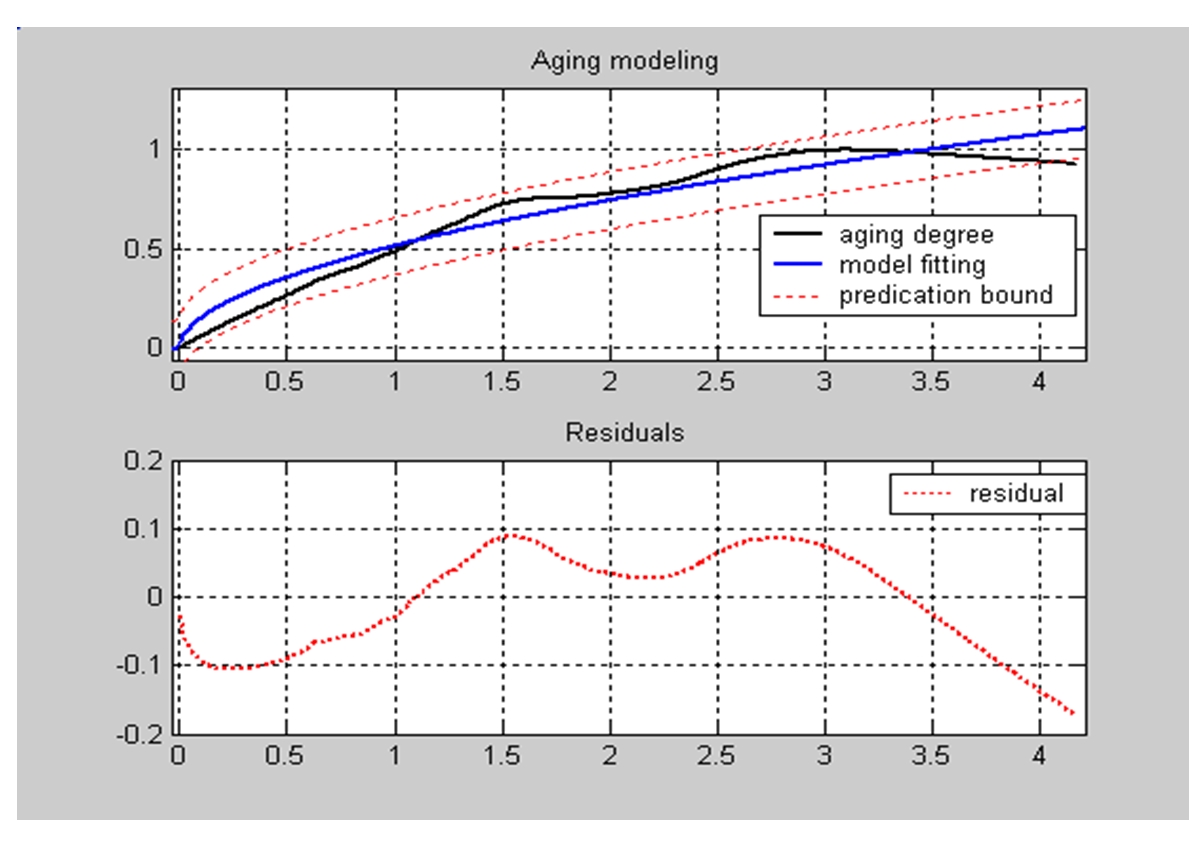}
\end{center}
\caption{
{\bf  The aging kinetic curve fitting and  model residual obtained by feedback loop model under workload $w_{1}$ in Helix server benchmark.}
}
\label{free_radical}
\end{figure}
\begin{figure}[!ht]
\begin{center}
\includegraphics[width=4in]{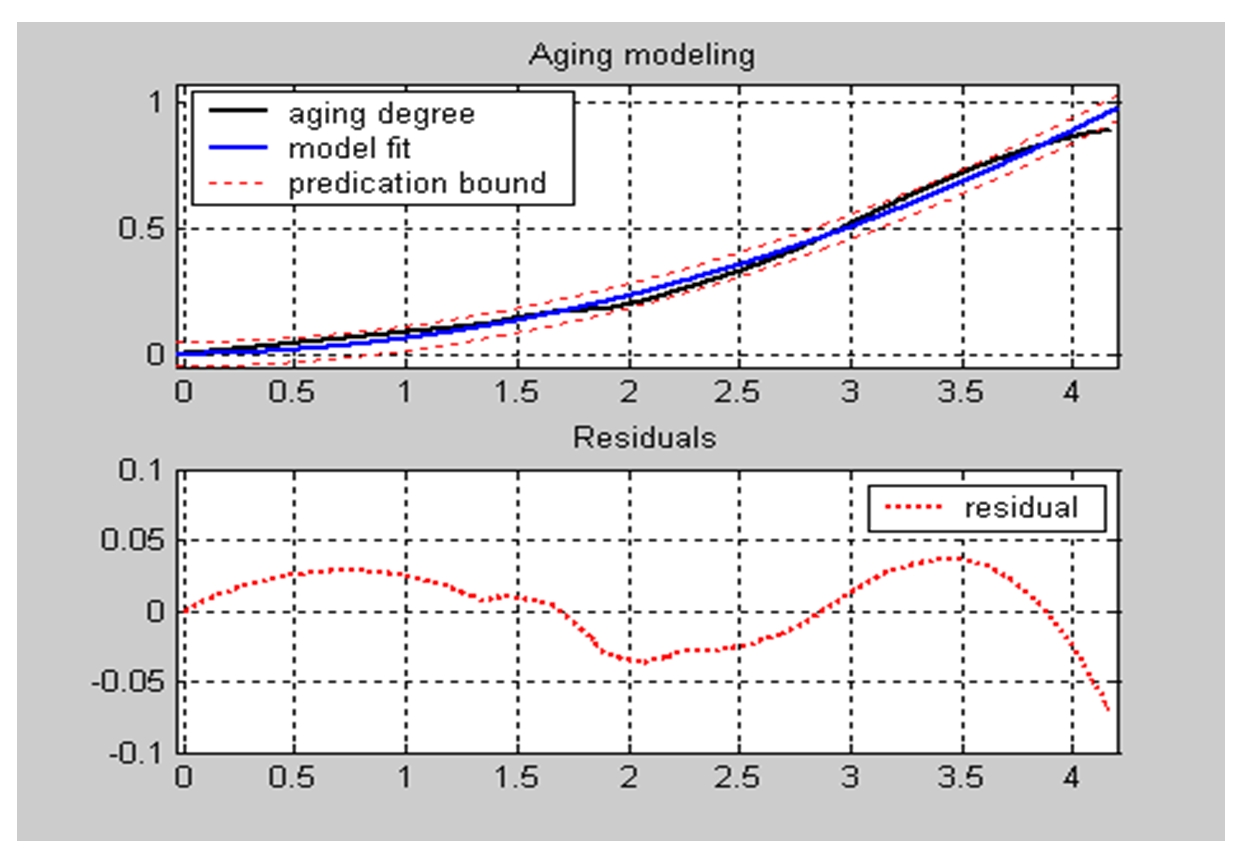}
\end{center}
\caption{
{\bf The aging kinetic curve fitting and  model residual obtained by feedback loop model under workload $w_{2}$ in Helix server benchmark. }
}
\label{free_radical}
\end{figure}
\begin{figure}[!ht]
\begin{center}
\includegraphics[width=4in]{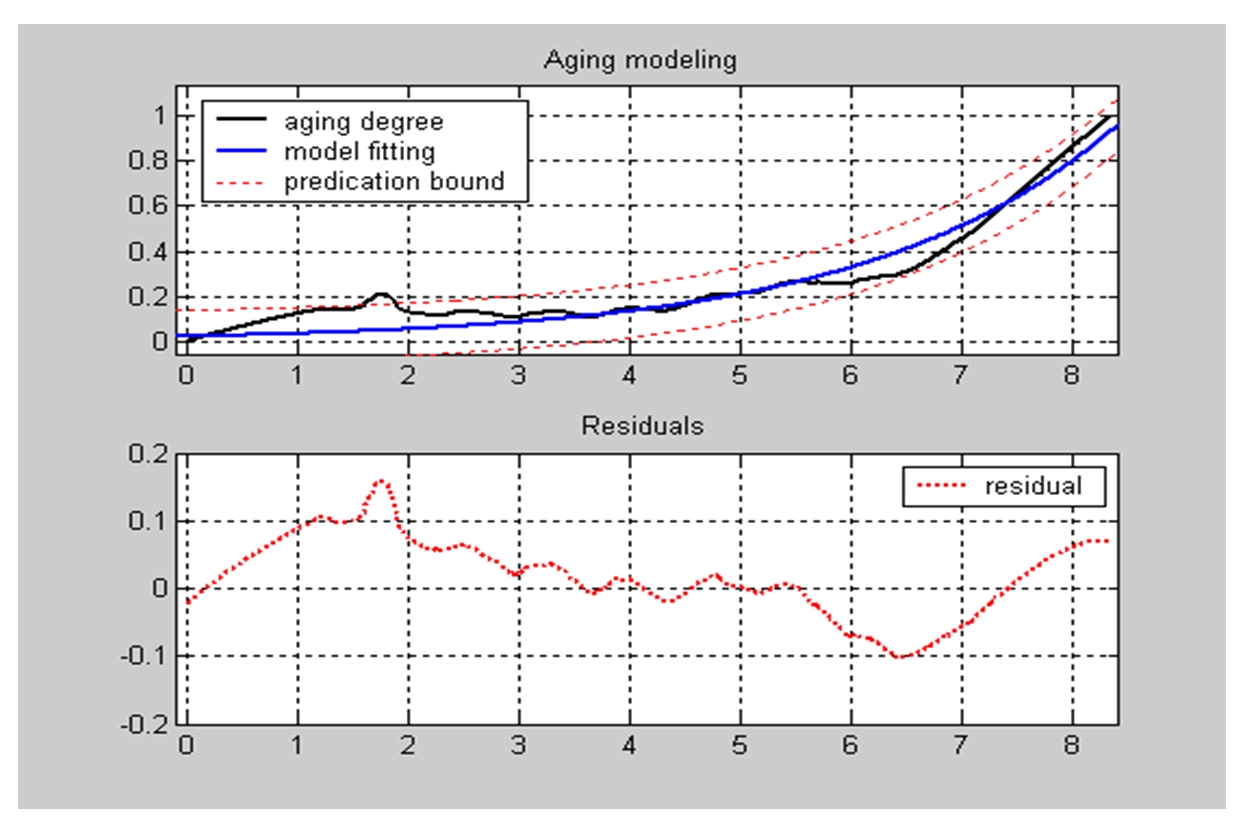}
\end{center}
\caption{
{\bf The aging kinetic curve fitting and  model residual obtained by feedback loop model under workload $w_{3}$ in Helix server benchmark.} 
}
\label{free_radical}
\end{figure}
\begin{figure}[!ht]
\begin{center}
\includegraphics[width=4in]{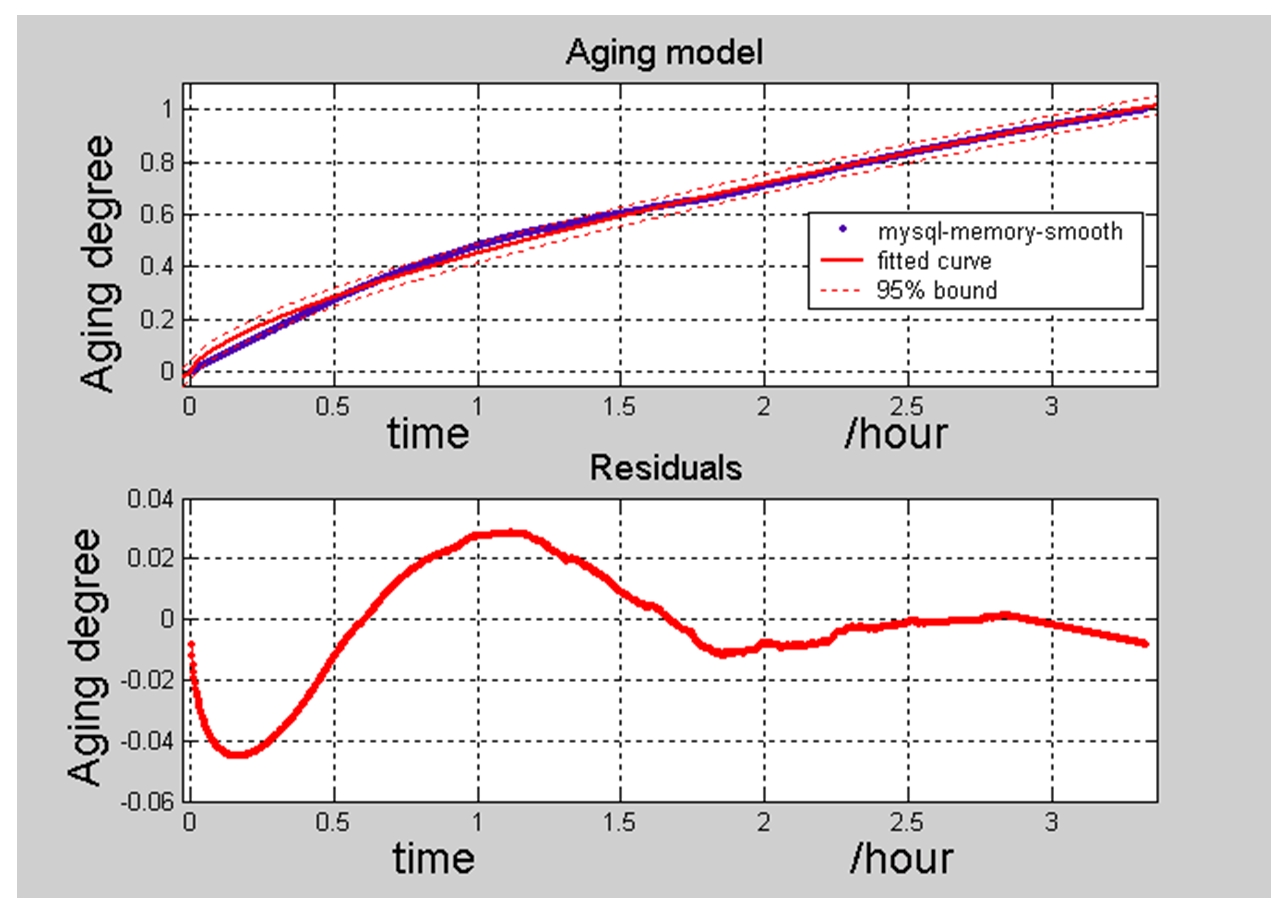}
\end{center}
\caption{
{\bf The aging kinetic curve fitting and  model residual obtained by feedback loop model in web server benchmark. } 
}
\label{free_radical}
\end{figure}
\begin{figure}[!ht]
\begin{center}
\includegraphics[width=4in]{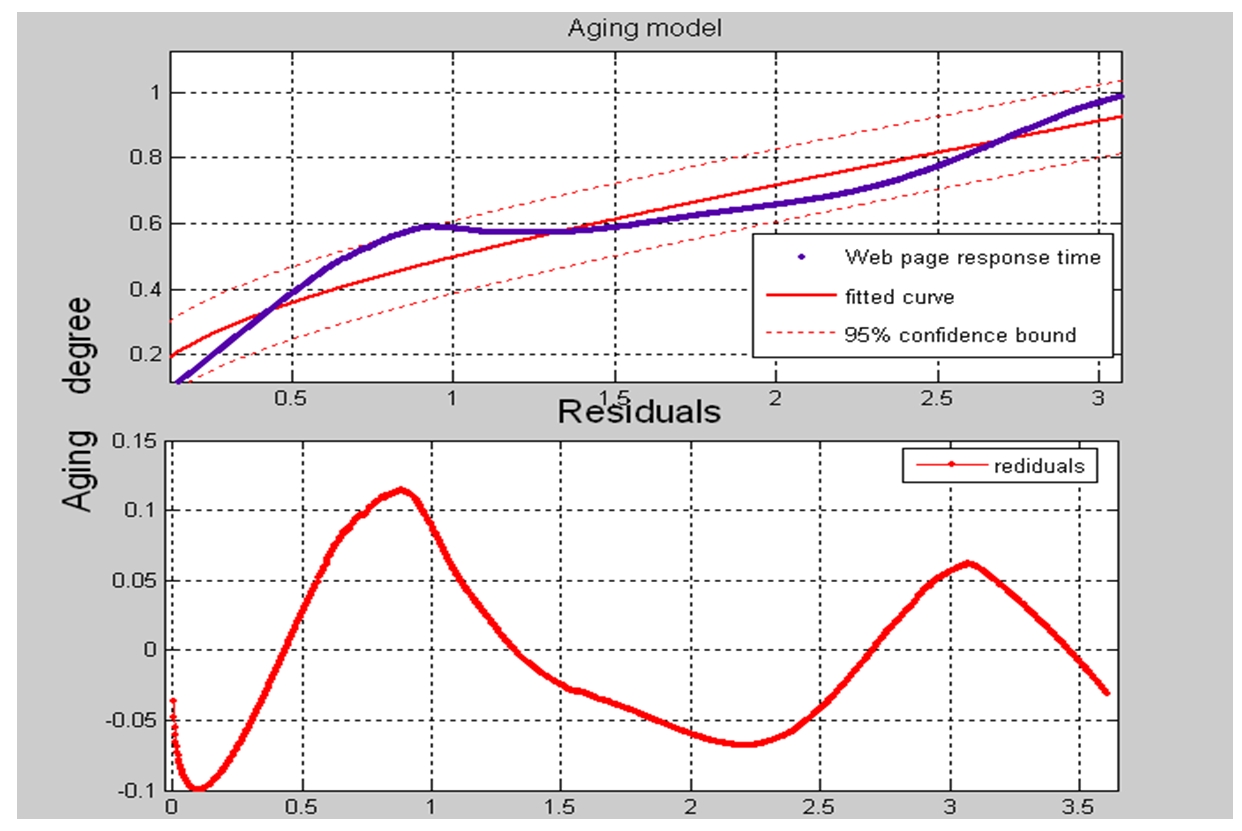}
\end{center}
\caption{
{\bf The aging kinetic curve fitting and  model residual obtained by feedback loop model in TPC-W benchmark. } 
}
\label{free_radical}
\end{figure}
\begin{figure}[!ht]
\begin{center}
\includegraphics[width=4in]{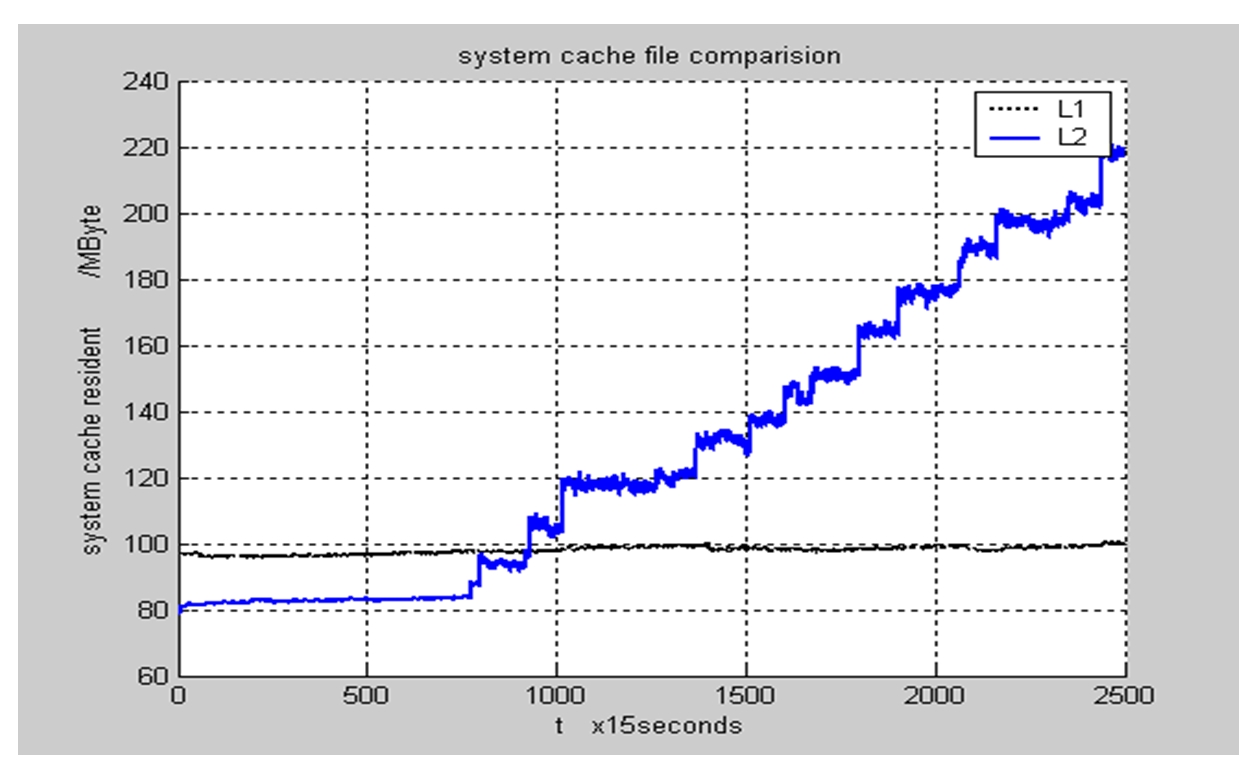}
\end{center}
\caption{
{\bf  The changes of cache content of Helix server under workload $L_{1}$ and $L_{2}$ }  
}
\label{free_radical}
\end{figure}
\begin{figure}[!ht]
\begin{center}
\includegraphics[width=4in]{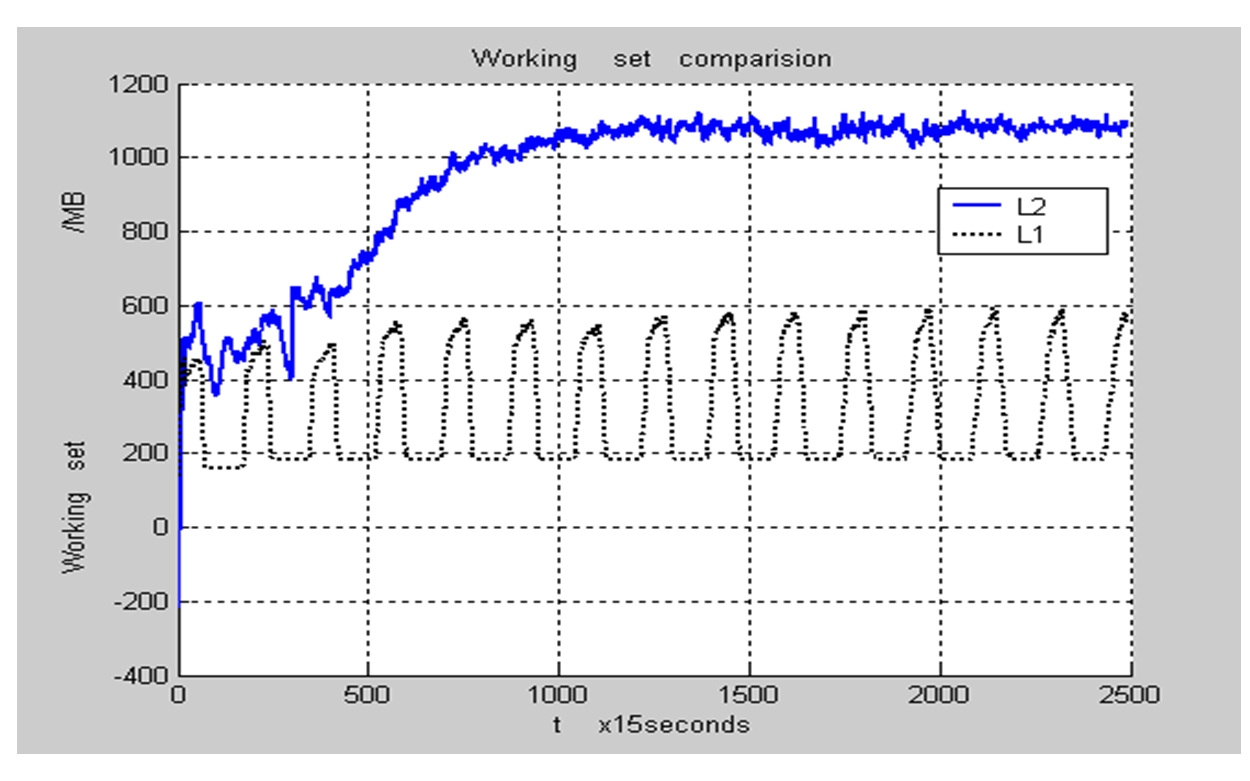}
\end{center}
\caption{
{\bf The changes of working set of Helix server under workload $L_{1}$ and $L_{2}$. }
}
\label{free_radical}
\end{figure}
\begin{figure}[!ht]
\begin{center}
\includegraphics[width=4in]{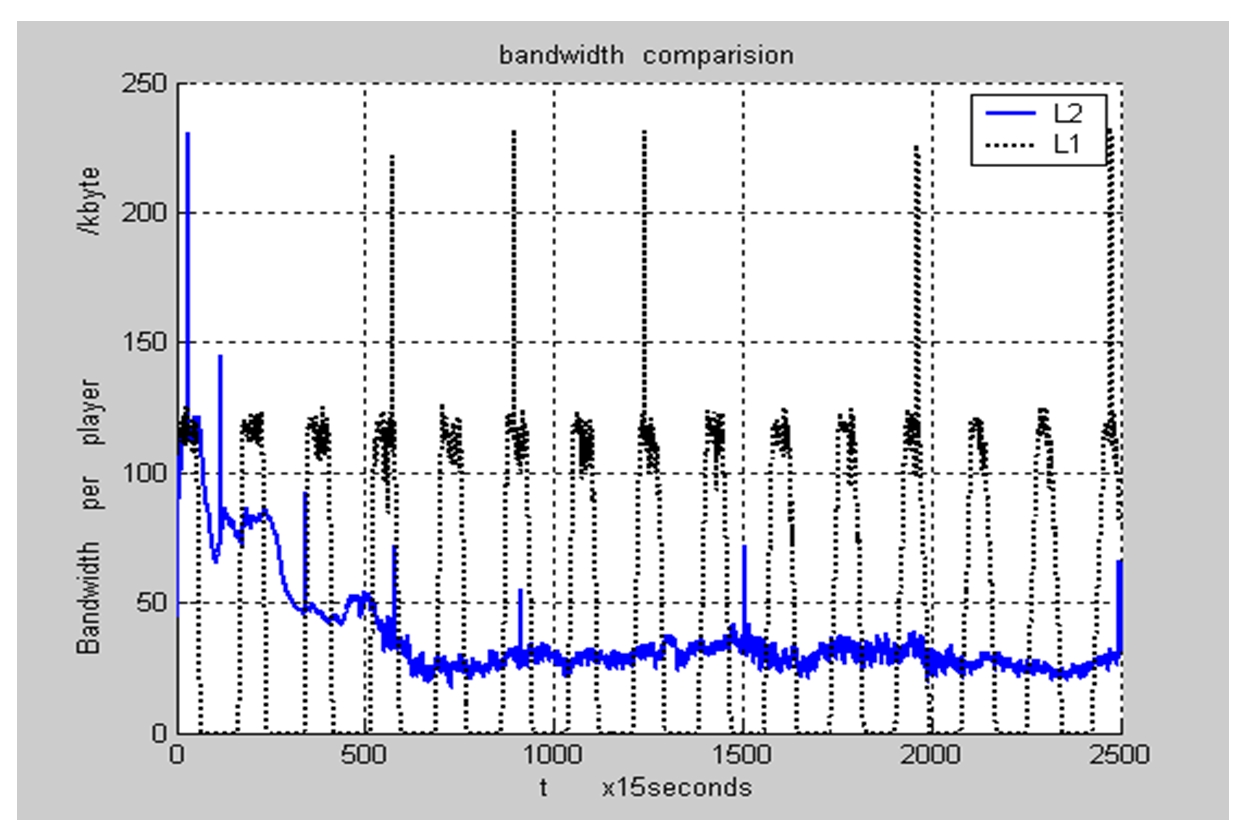}
\end{center}
\caption{
{\bf The changes of working set of Helix server under workload $L_{1}$ and $L_{2}$. }  
}
\label{free_radical}
\end{figure}
\begin{figure}[!ht]
\begin{center}
\includegraphics[width=4in]{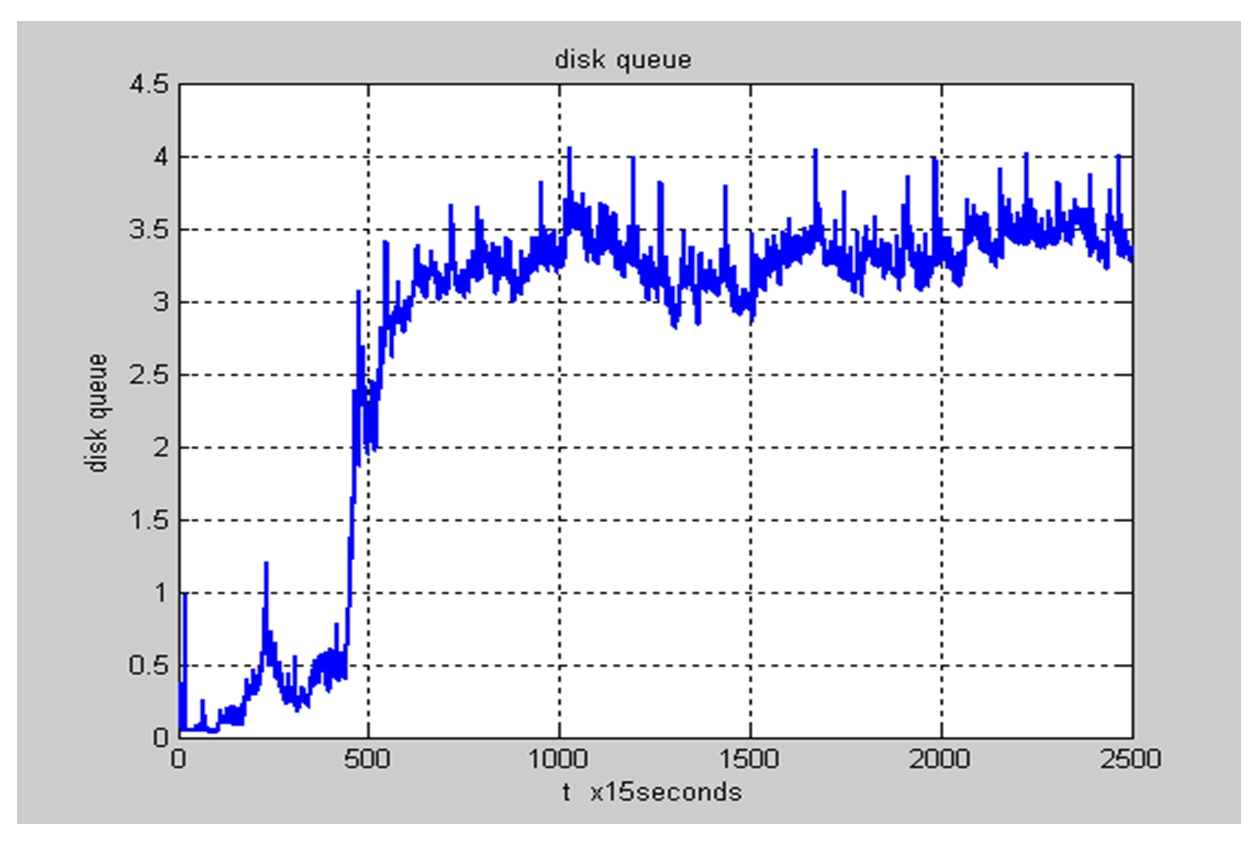}
\end{center}
\caption{
{\bf The sharp increasing of disk queue length in Helix server benchmark. } 
}
\label{free_radical}
\end{figure}
\begin{figure}[!ht]
\begin{center}
\includegraphics[width=6in]{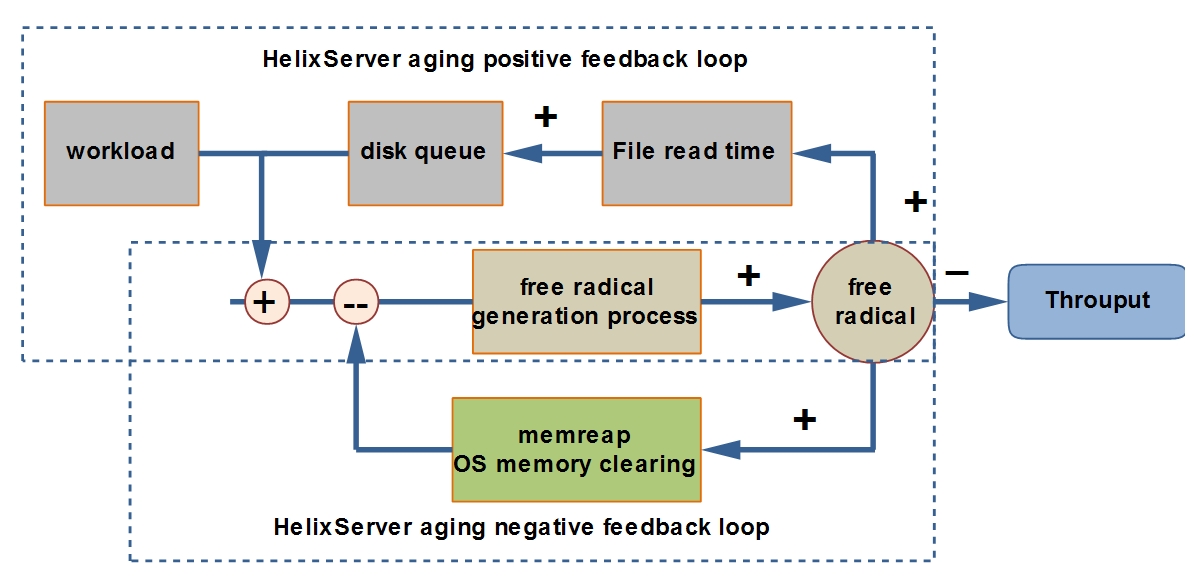}
\end{center}
\caption{
{\bf The negative and positive feedback loops in Helix server benchmark.}  
}
\label{free_radical}
\end{figure}
\begin{figure}[!ht]
\begin{center}
\includegraphics[width=4in]{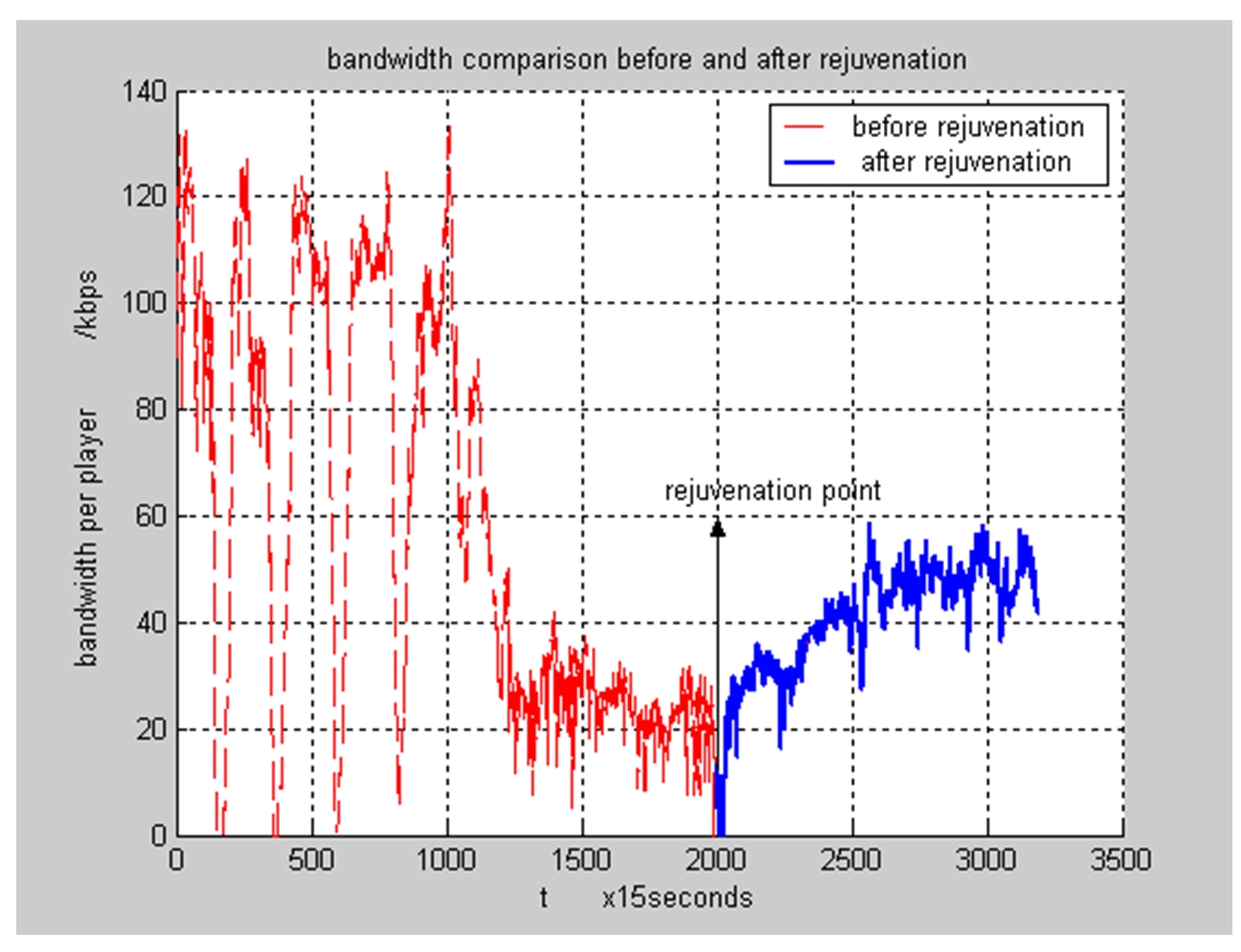}
\end{center}
\caption{
{\bf The changes of `bandwidth per player' using the first workload regulation method. } 
}
\label{free_radical}
\end{figure}
\begin{figure}[!ht]
\begin{center}
\includegraphics[width=4in]{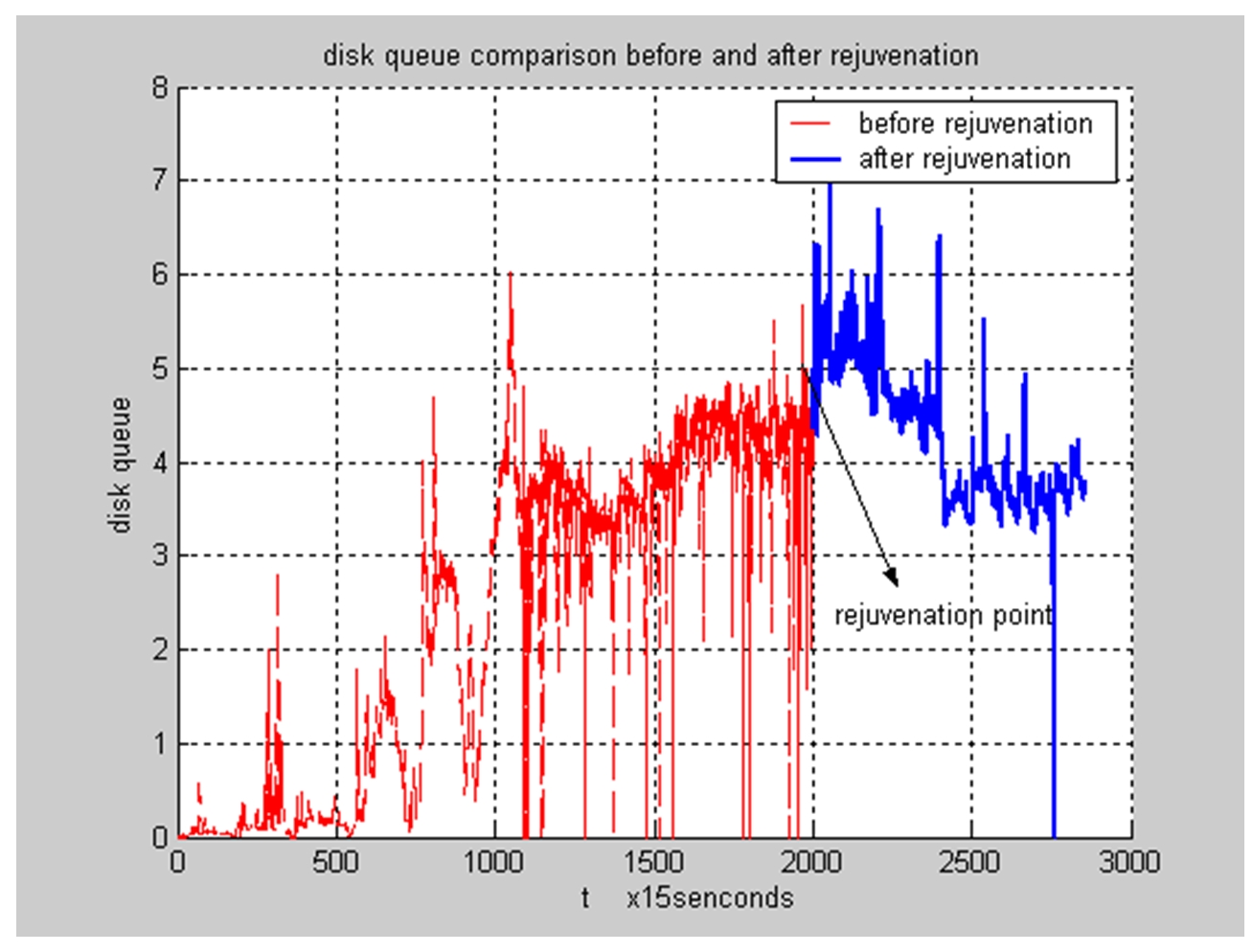}
\end{center}
\caption{
{\bf The changes of `disk queue length' using the first workload regulation method. } 
}
\label{free_radical}
\end{figure}
\clearpage
\begin{figure}[!ht]
\begin{center}
\includegraphics[width=4in]{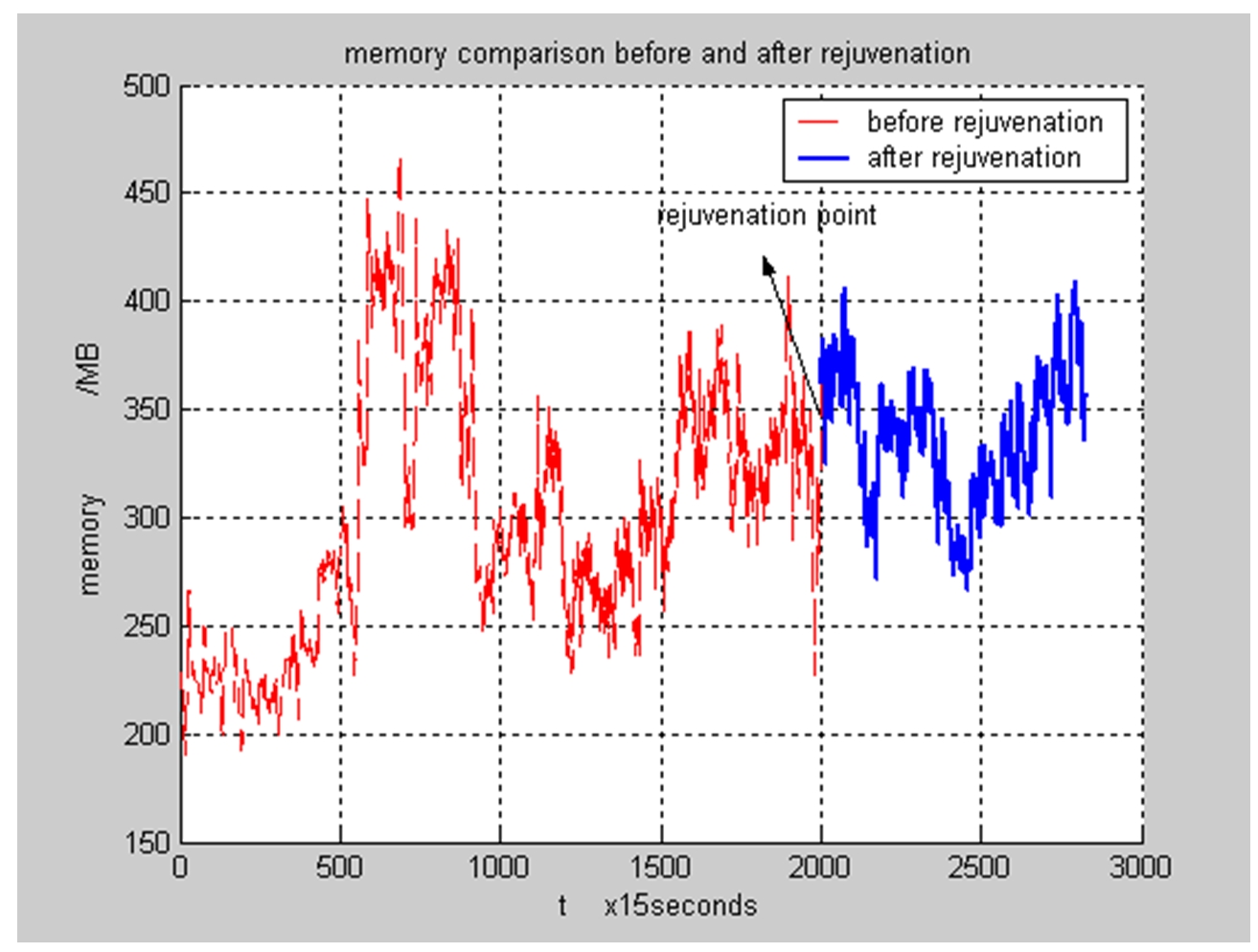}
\end{center}
\caption{
{\bf The changes of `memory utilization' using the first workload regulation method. } 
}
\label{free_radical}
\end{figure}
\begin{figure}[!ht]
\begin{center}
\includegraphics[width=4in]{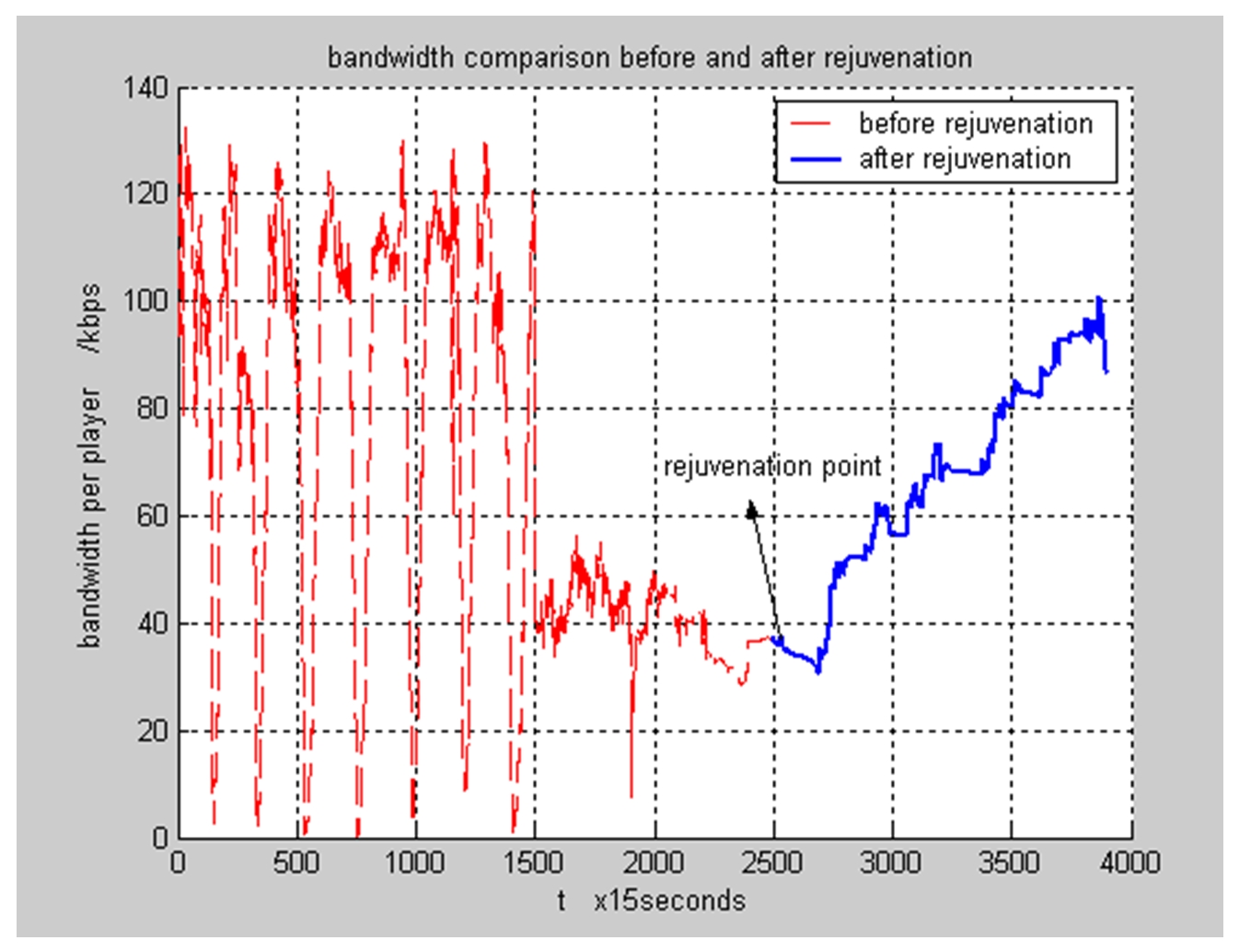}
\end{center}
\caption{
{\bf The changes of `bandwidth per player' using the second workload regulation method. }  
}
\label{free_radical}
\end{figure}
\begin{figure}[!ht]
\begin{center}
\includegraphics[width=4in]{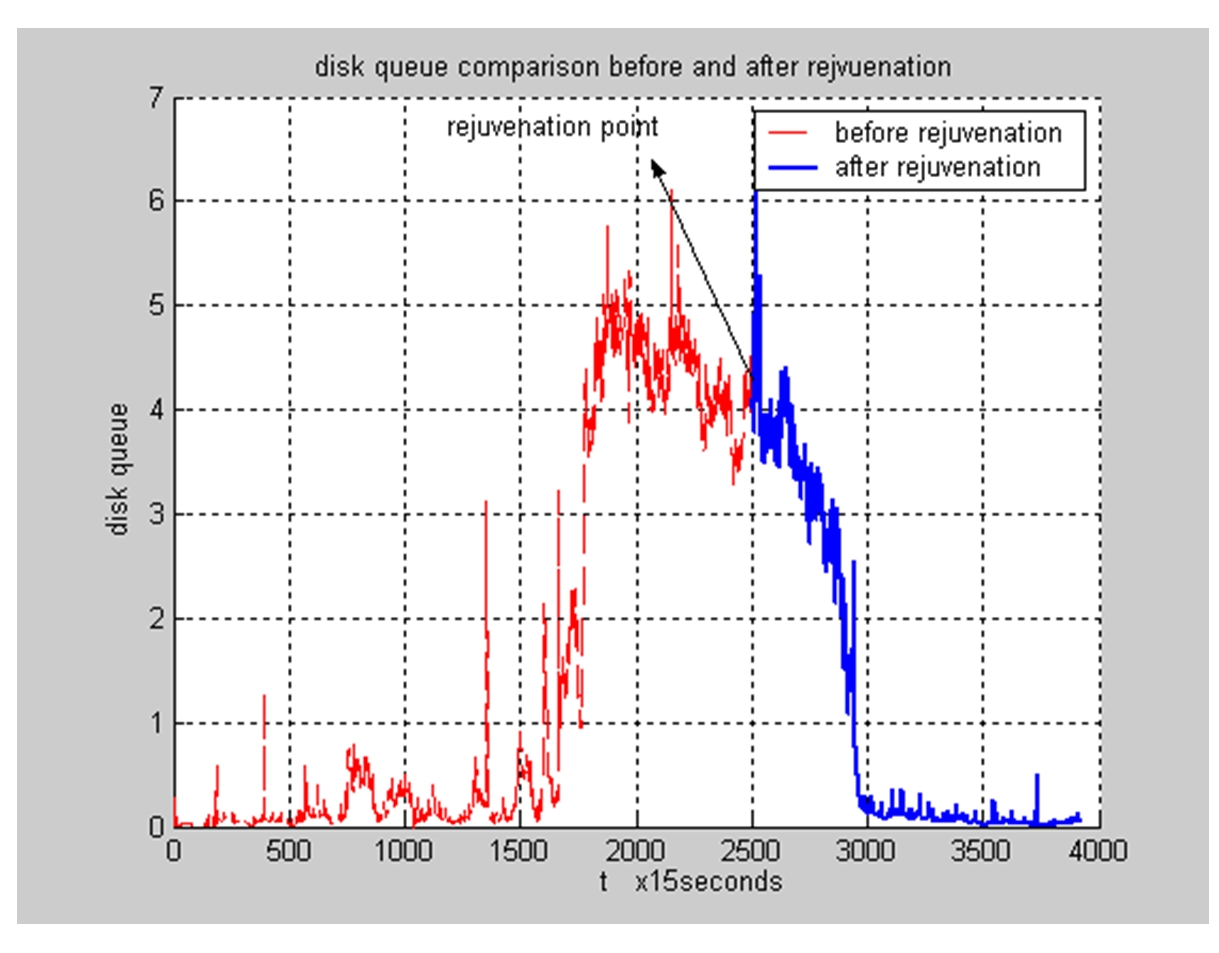}
\end{center}
\caption{
{\bf The changes of `disk queue length' using the second workload regulation method. } 
}
\label{free_radical}
\end{figure}
\begin{figure}[!ht]
\begin{center}
\includegraphics[width=4in]{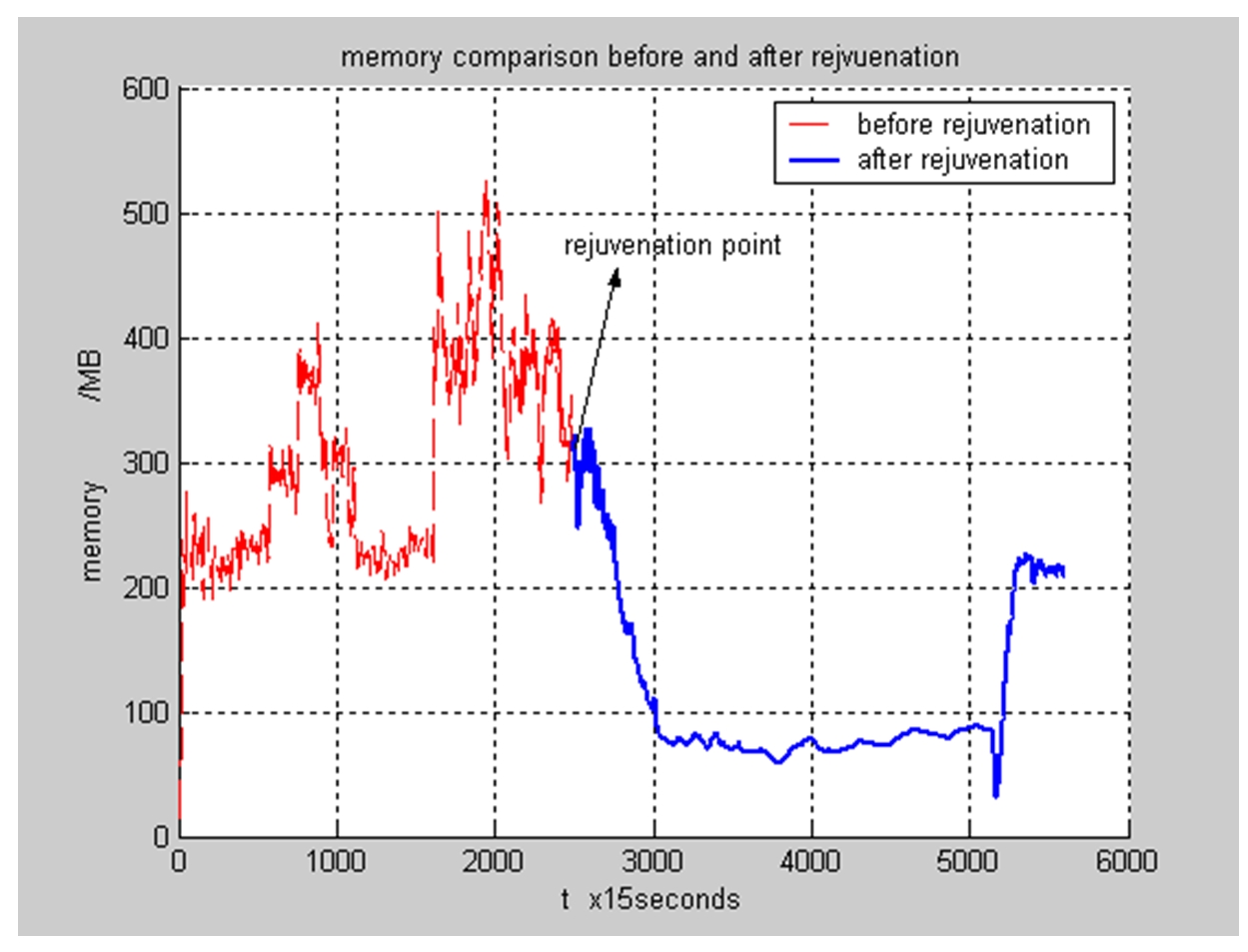}
\end{center}
\caption{
{\bf The changes of `memory utilization' using the second workload regulation method.  }  
}
\label{free_radical}
\end{figure}
\begin{figure}[!ht]
\begin{center}
\includegraphics[width=4in]{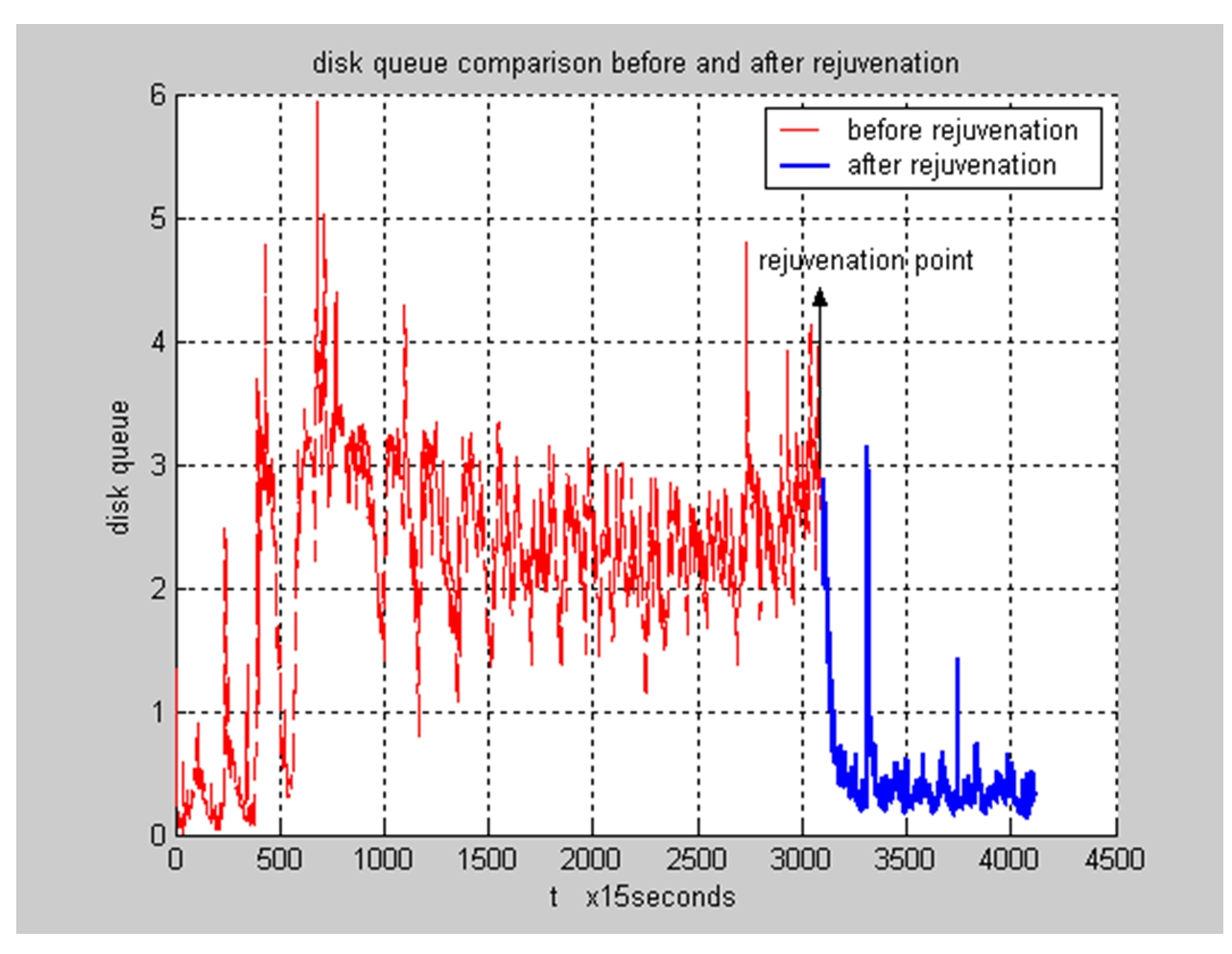}
\end{center}
\caption{
{\bf The changes of `disk queue length' using the disk queue regulation method.  }  
}
\label{free_radical}
\end{figure}
\begin{figure}[!ht]
\begin{center}
\includegraphics[width=4in]{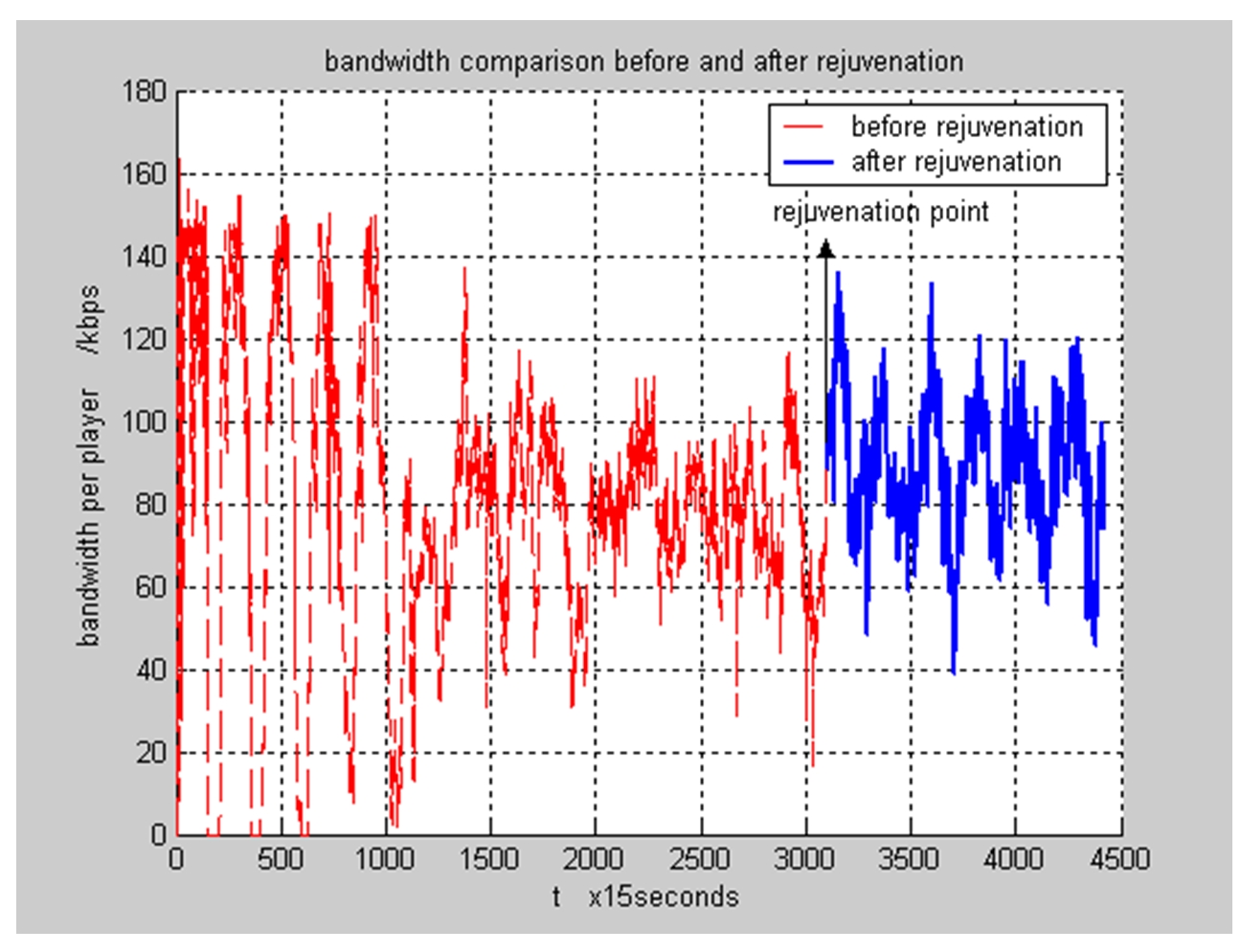}
\end{center}
\caption{
{\bf The changes of `bandwidth per player' using the disk queue regulation method. }  
}
\label{free_radical}
\end{figure}
\begin{figure}[!ht]
\begin{center}
\includegraphics[width=4in]{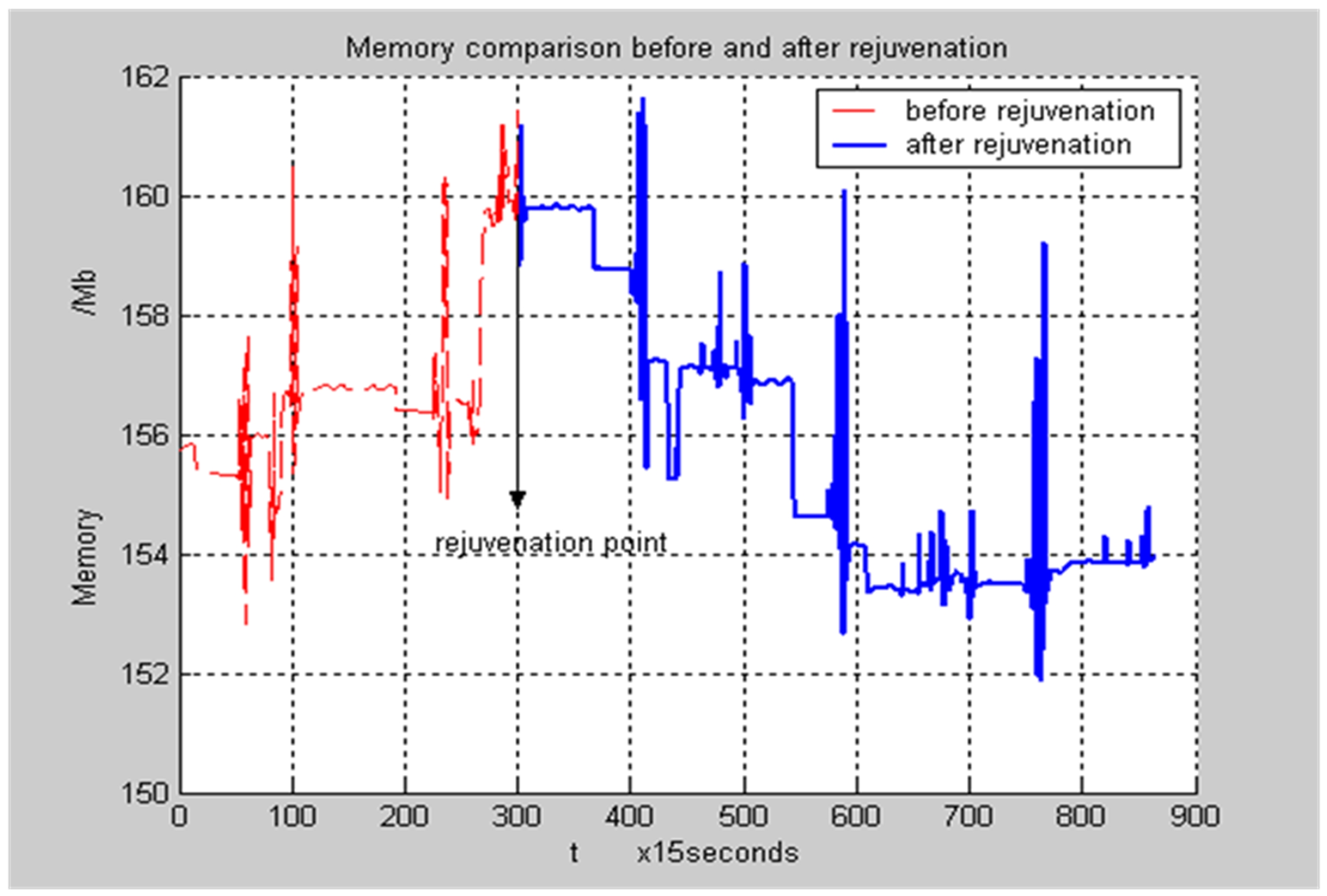}
\end{center}
\caption{
{\bf The changes of `memory utilization' using the memory cleaning method.  }  
}
\label{free_radical}
\end{figure}
\begin{figure}[!ht]
\begin{center}
\includegraphics[width=4in]{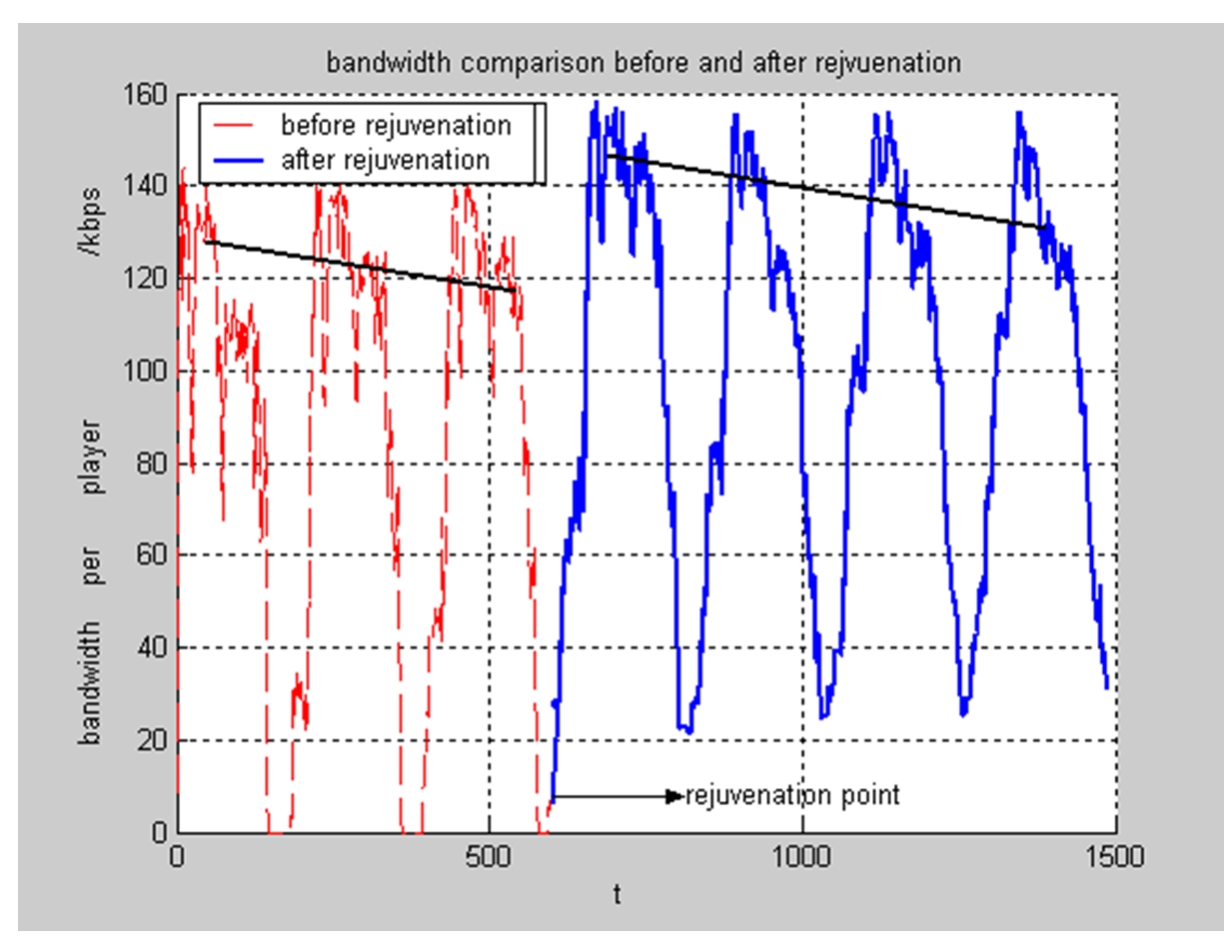}
\end{center}
\caption{
{\bf The changes of `bandwidth per player' using the memory cleaning method.   }  
}
\label{free_radical}
\end{figure}
\begin{figure}[!ht]
\begin{center}
\includegraphics[width=4in]{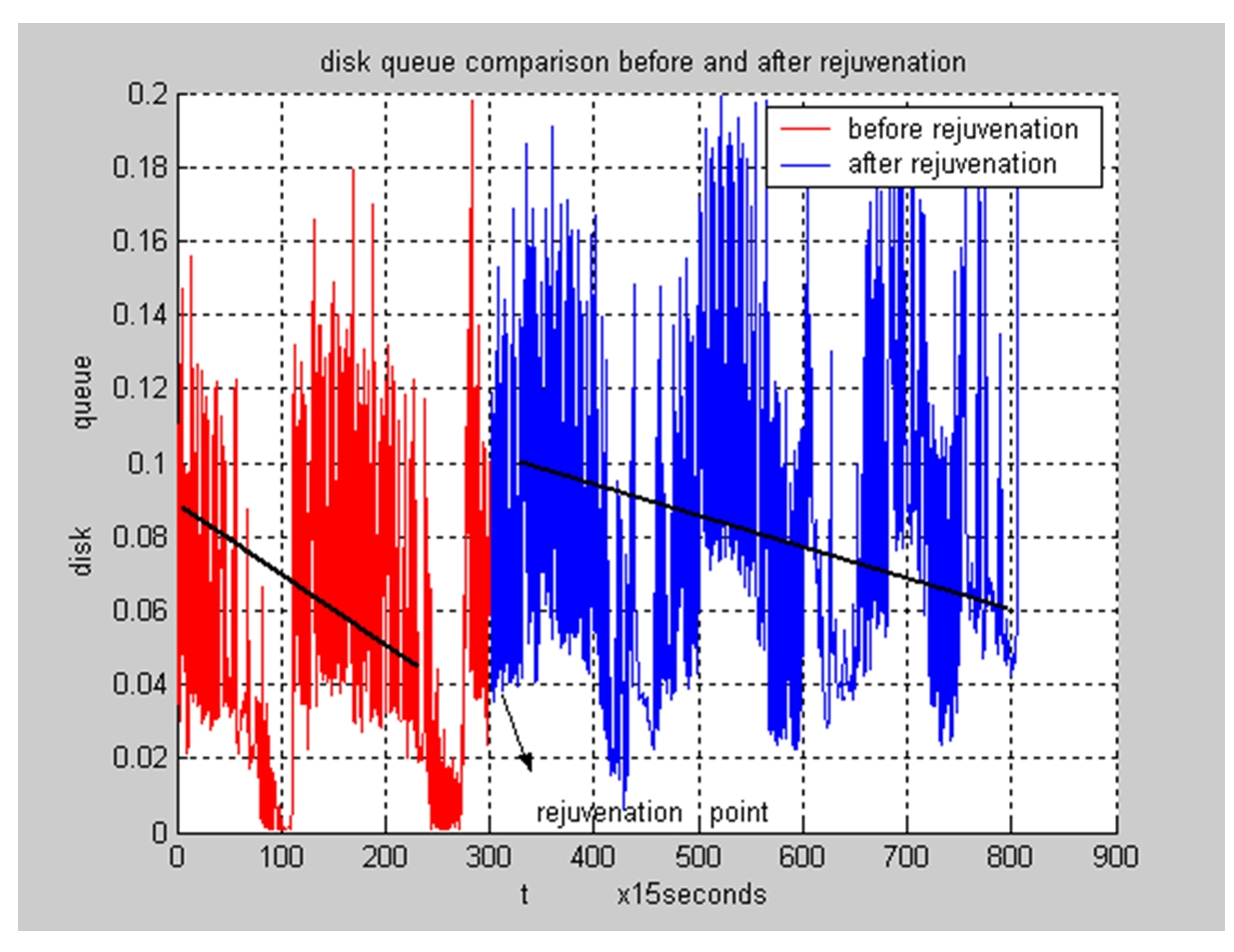}
\end{center}
\caption{
{\bf The changes of `disk queue length' using the memory cleaning method. }  
}
\label{free_radical}
\end{figure}
\begin{figure}[!ht]
\begin{center}
\includegraphics[width=4in]{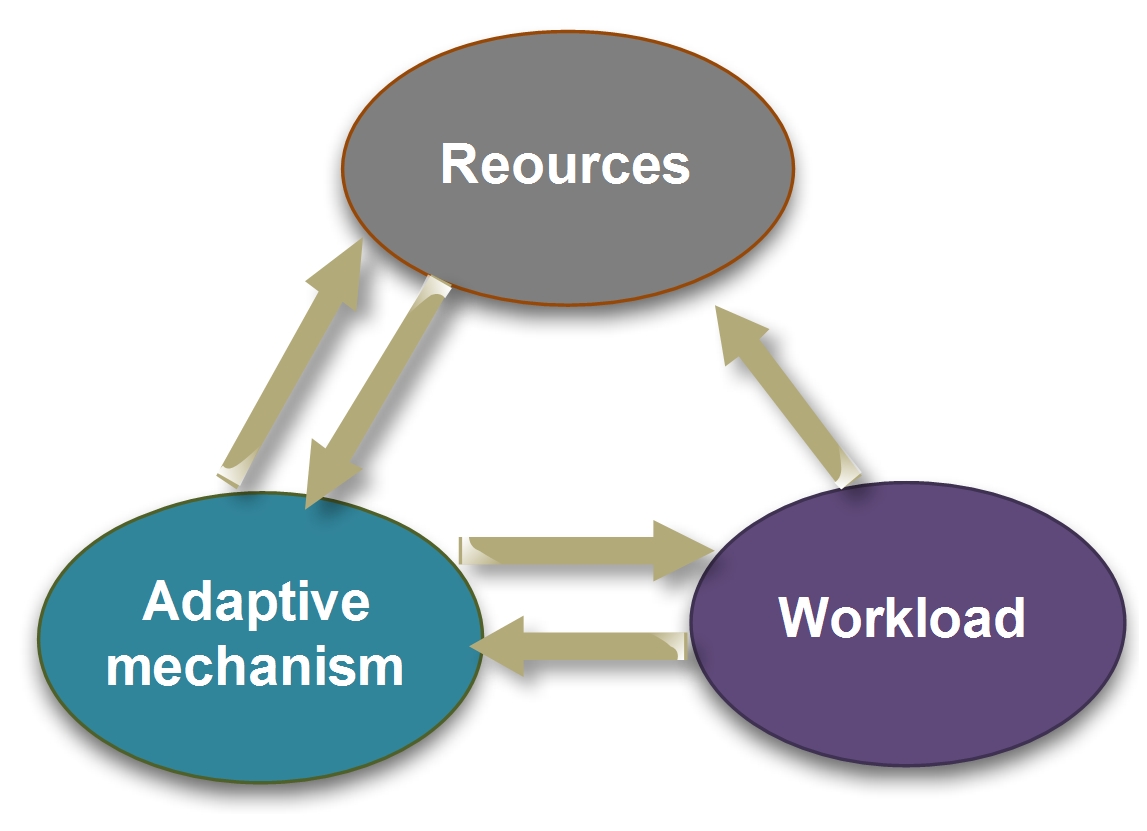}
\end{center}
\caption{
{\bf The interplay among the three aging impact factors :\textit{resource, adaptive mechanism, workload} }  
}
\label{free_radical}
\end{figure}
\begin{figure}[!ht]
\begin{center}
\includegraphics[width=6in]{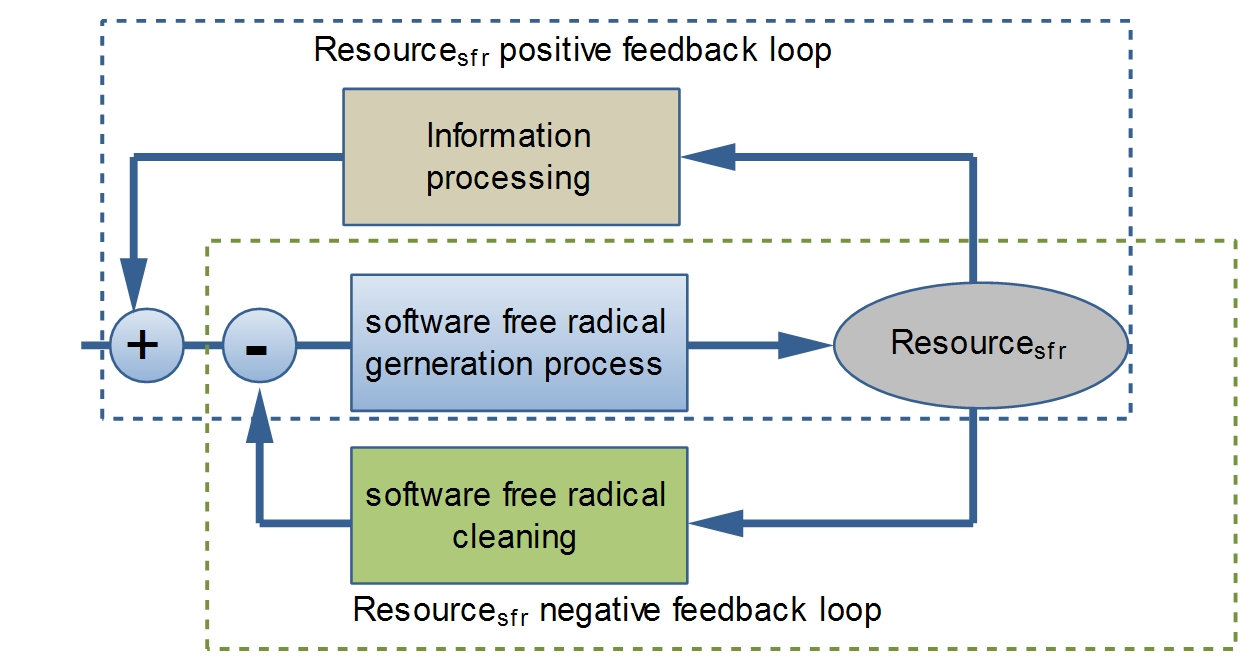}
\end{center}
\caption{
{\bf The integrated negative and positive feedback loops in software systems. }  
}
\label{free_radical}
\end{figure}
\begin{figure}[!ht]
\begin{center}
\includegraphics[width=4in]{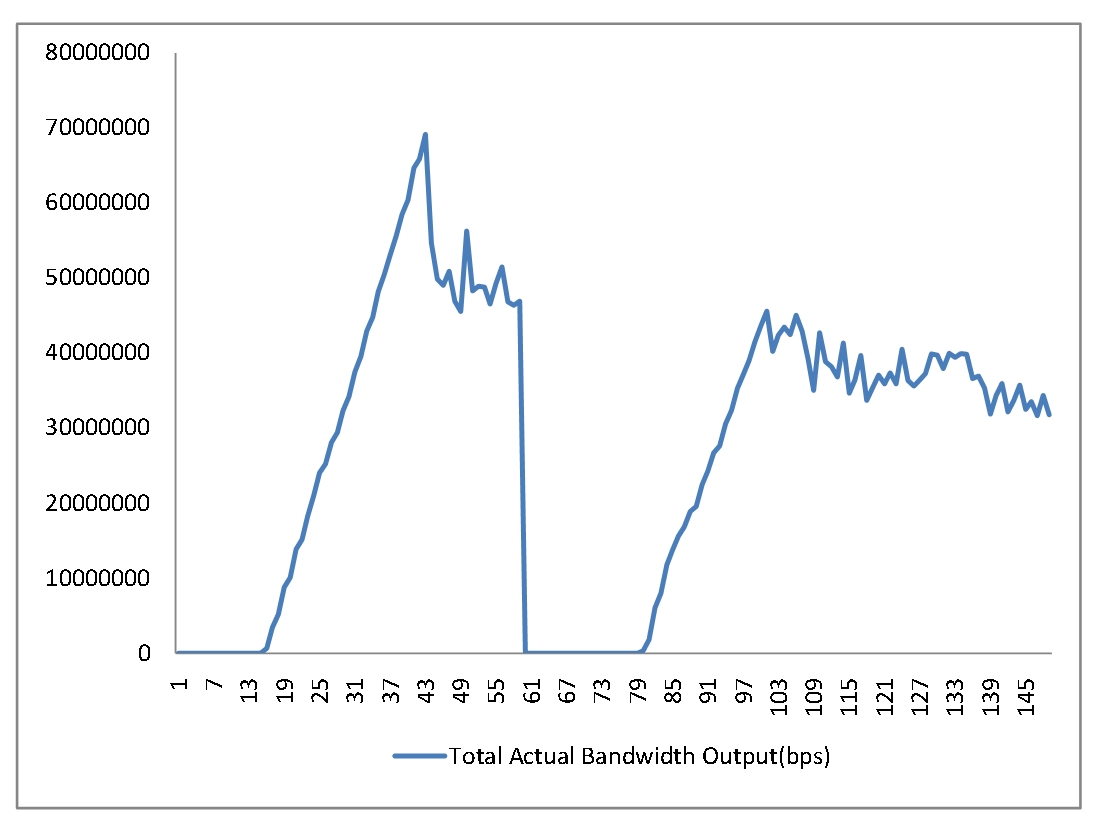}
\end{center}
\caption{
{\bf The capacity test in Helix server benchmark. } 
}
\label{free_radical}
\end{figure}
\clearpage
\section*{Tables}
\begin{table}[!ht]
\caption{
\bf{The estimated model parameters and the model errors}}
\begin{center}
\begin{tabular}{|c|r|c|c|c|c|c|}
\hline
\multicolumn{ 2}{|c|}{Software} &          K &          $\alpha$ &          $\beta$ &       RMSE &   R-square \\
\hline
\multicolumn{ 1}{|c|}{Helix server} &         w1 &     0.5133 &   7.02E-11 &     0.5357 &    0.07355 &     0.9361 \\
\hline
\multicolumn{ 1}{|c|}{} &         w2 &       0.06 &     0.0294 &      1.858 &     0.0247 &     0.9924 \\
\hline
\multicolumn{ 1}{|c|}{} &         w3 &      0.225 &      0.446 &   2.29E-14 &     0.0592 &      0.931 \\
\hline
\multicolumn{ 2}{|c|}{TPC-W} &     0.4504 &   1.09E-09 &     0.6693 &    0.01817 &     0.9955 \\
\hline
\multicolumn{ 2}{|c|}{Web server} &     0.4638 &     0.0676 &       0.43 &    0.05639 &     0.9495 \\
\hline
\end{tabular}  
\end{center}
\end{table}
\begin{table}[!ht]
\caption{
\bf{The attributes of workload in Helix server benchmark}}
\begin{center}
\begin{tabular}{|c|c|}
\hline
 Attribute &    Meaning \\
\hline
client\_count & Total emulated Helix clients  \\
\hline
\multicolumn{ 1}{|c|}{file\_ploy} & Access distribution: 0 denotes random access, 1 denotes sequential   \\

\multicolumn{ 1}{|c|}{} & access, 2 denotes Poisson access, 3 denotes single file access  \\
\hline
\multicolumn{ 1}{|c|}{file\_object} & Files allowed accessed, users have no authorization to access 
 \\

\multicolumn{ 1}{|c|}{} & filesbeyond  these files. \\
\hline
file\_max\_object  & Maximal files allowed accessed within 0 to file\_object concurrently \\
\hline
sleep\_time & Interval between two continuous requests \\
\hline
file\_difference & Whether requested files are the same: 0 denotes different, 1 denotes the same \\
\hline
\end{tabular}  
\end{center}
\end{table}
\end{document}